\documentclass{birkjour}

 \theoremstyle{definition}
 
 \theoremstyle{remark}

 \numberwithin{equation}{section}

\begin{document}

\title[]
 {Emmy Noether on Energy Conservation\\ in General Relativity}

\author[]{David E. Rowe\\Mainz University}




\maketitle

\noindent
Abstract: During the First World War, the status of energy conservation in general relativity was one of the most hotly debated questions surrounding Einstein's new theory of gravitation. His approach to this aspect of general relativity differed sharply from another set forth by Hilbert, even though the latter conjectured in 1916 that both theories were probably equivalent. Rather than pursue this question himself, Hilbert chose to 
charge Emmy Noether with the task of probing  the mathematical foundations of these two theories. Indirect references to her results came out two years later when Klein began to examine this question again with Noether's assistance. Over several months,  Klein and Einstein pursued these matters in a lengthy correspondence, which culminated with several publications, including Noether's now famous paper ``Invariante Variationsprobleme'' \cite{Noether1918b}. The  present account focuses on the earlier discussions from 1916 involving Einstein, Hilbert, and Noether. In these years, a Swiss student named R.J. Humm was studying relativity in G\"ottingen, during which time he transcribed part of Noether's lost manuscript on Hilbert's invariant energy vector. By making use of this 9-page manuscript, it is possible to reconstruct
 the arguments Noether set forth in response to Hilbert's conjecture. Her results turn out to be closely related to the findings Klein published two years later, thereby highlighting, once again, how her work significantly deepened contemporary  understanding of the mathematical underpinnings of general relativity.

\section{Introduction}

Emmy Noether's paper ``Invariante Variationsprobleme'' \cite{Noether1918b}  is regarded today as one of her most important works, especially in view of its relevance for mathematical physics \cite{Uhl}. Those familiar with her many other achievements might wonder why these have largely  been cast in the shadows by Noether's Theorem, the famous result accounting for the relationship between symmetries in physical systems and their related conservation laws.\footnote{For a detailed analysis of \cite{Noether1918b} and its slow reception by the mathematical community, see \cite{Kosmann-Schwarzbach2006}. See also the commentary in \cite{Siegmund-Schultze2011}.}
 To be sure, standard accounts of Emmy  Noether's life have never claimed that her 1918 paper was particularly significant, and for good reason. Her 
 influence and eventual fame as a mathematician had virtually nothing to do with physics or the calculus of variations; these  stemmed instead from her contributions to modern algebra.\footnote{See \cite{Alex}, \cite{Weyl-3}, \cite{Dick}, \cite{Kim}, and \cite{Kor}.}
 Considering these circumstances, it is natural to ask what motivated Noether to take up this topic in the first place. \cite{Rowe1999} deals with  how Noether's paper arose from discussions in G\"ottingen concerning the status of  energy conservation laws in general relativity. This paper focusses on an earlier discussion that
 arose in 1916 after Albert Einstein and David Hilbert published their first papers addressing the role of energy conservation in general relativity.

 As will be shown below, the approaches taken by Einstein and Hilbert to this aspect of the theory 
 differed strikingly. Hilbert's  short paper \cite{Hilbert1915} was written in great haste and afterward substantially revised when he read the page proofs. Its contents 
baffled many readers, including Einstein. In 1918, Emmy Noether was working closely with Felix Klein, who was determined  to decipher the mathematical meaning of Hilbert's invariant energy vector. Allusions to Noether's role in earlier discussions with Hilbert can be found in Klein's published papers, \cite{Kl-1}  and \cite{Kl-2}. 
Little has been written, however, about what Noether contributed to these conversations from 1916, mainly due to lack of documentary evidence that might shed more  light on her activities during the war years. 
It is my hope that the present paper will help to clarify an important episode in this story. Here I will mainly focus on her efforts, starting in early 1916, to assist Hilbert's researches on general relativity, while touching only briefly on her subsequent work with Klein, which culminated with the publication of \cite{Noether1918b}.

In the course of exploring the foundations of his general theory of relativity, Einstein had experimented with variational methods \cite{Einstein1914}. Hilbert was the first, however, to use a variational principle to derive fully covariant gravitational field equations in the form of Euler-Lagrange equations. He
published this result in the first of his two papers on ``The Foundations of Physics''
 \cite{Hilbert1915}.  There he emphasized that the resulting system of 14 field equations was {\it not} independent; instead it 
satisfied four identical relations, which he interpreted as establishing a linkage between gravity and electromagnetism. However, the precise nature of these relations, and in particular their physical significance, remained obscure up until the publication of \cite{Noether1918b}, which completely clarified this question. 

More mysterious still was what Hilbert called his invariant energy equation, which he based on a complicated construct that came to be known as Hilbert's energy vector $e^l$.\footnote{As noted below, its definition
 (\ref{eq:Hilbert-vector})  depends linearly on an arbitrary infinitesimal vector $p^l$, so $e^l$ should be conceived as a vector field. For a detailed analysis of Hilbert's approach to conservation of energy-momentum, both in \cite{Hilbert1915} and in the earlier unpublished version in \cite{Hilbert2009}, 
see \cite{Renn/Stachel2007}.}
 He derived this vector using classical techniques for producing differential invariants, an approach that differed sharply from Einstein's much more physically motivated derivation of energy conservation in \cite{Ein-6}. Klein later  showed how Hilbert's energy vector arose naturally from the variational framework used in his theory \cite{Kl-2}. Six years later, when Hilbert decided to publish a modified account of his original theory in  \cite{Hilbert1924}, he dropped all reference to his earlier approach to energy conservation, a clear indication that he no longer felt it had any importance for  his unified field theory. Already in January 1918 Klein had exposed various hasty claims made in  \cite{Hilbert1915}. His critique in  \cite{Kl-1} set the stage for Noether's insightful analysis that showed precisely how conservations laws and certain identities based on them arise in theories based on a variational principle. Klein took it upon himself to analyze the various
proposals for conservation laws in
differential form in \cite{Kl-2}. In the course of doing so, he gave a simplified
and much clearer
derivation of Hilbert's invariant energy equation (\ref{eq:Hilbert-energy}). He also succeeded in characterizing Einstein's formulation of energy conservation as presented in \cite{Ein-8}. Soon afterward, Klein took up Einstein's integral form for energy-momentum conservation in \cite{Kl-7}.\footnote{For a summary account of these issues as seen three years later, see \cite[175--178]{Pau-2}.} In all  of these studies he was assisted by Emmy Noether.

When 
 Einstein began to study Hilbert's paper \cite{Hilbert1915}  in earnest in May 1916, he naturally wondered whether there might be some deep connection between his own findings and Hilbert's energy equation. Hilbert thought this was probably the case, and he wrote as such to Einstein. He also informed him that he had already asked Emmy Noether to investigate this question, a circumstance that suggests he may have been disinclined to pursue this matter himself. 
What transpired afterward remains somewhat shrouded in mystery, but the present account will show that already by 1916 Noether had taken an important step toward solving this problem. In that year, she discovered that 
 Hilbert's energy theorem as well as Einstein's formula for energy conservation shared a formal property closely connected with the field equations for gravitation. Although she never published on this topic, direct allusions to her discovery came to the surface in 
 early 1918 when Klein and Hilbert published  \cite{Kl-1}, an exchange of letters  concerning the status of energy conservation in general relativity.  Thanks to 
 the recovery of a 9-page manuscript based on Noether's work, we can now reconstruct in outline 
 the arguments she set forth in response to Hilbert's inquiry.
 In  recounting this story, I have shifted the focus away from the immediate events of  1918 that led to 
 Noether's seminal achievement, her paper \cite{Noether1918b}. In the course of its telling, however, it will become clear that the earlier events from 1916 -- in particular Noether's findings with respect to energy conservation in the theories of Hilbert and Einstein -- directly presaged her later work, which arose from Klein's determination to clarify these issues.

\section{On the Research Agendas of Hilbert and Klein}

In the late
spring of 1915, only shortly after Emmy Noether's arrival, Einstein came to G\"ottingen to deliver a series of six two-hour
lectures on his new theory of gravitation, the general theory of 
relativity.\footnote{Einstein had been invited to G\"ottingen once before by Hilbert, in 1912,
but declined that invitation (Albert Einstein to
David Hilbert, 4 October 1912, in \cite[321--322]{Ein-3}.}
Einstein was pleased with the reception he was accorded, and
expressed particular pleasure with Hilbert's reaction. ``I am very
enthusiastic about Hilbert,'' he wrote Arnold Sommerfeld, ``an important
man!'' \cite[147]{Einstein1998a}.
 Hilbert had a long--standing interest in mathematical physics
 (see \cite{Corry2004}, \cite{Corry2007}). 
 Following Hermann Minkowski's lead, he and other G\"ottingen mathematicians felt 
 strongly drawn to the formal elegance of relativity theory. For Hilbert, who had been advocating an axiomatic approach to physics for many years, relativity was ready-made for this program. In later years, he liked to joke that ``physics had become too difficult
for the physicists,'' a quip that was probably not intended all too seriously (\cite[347]{Weyl-7}).
Although little is known about what transpired during the week of his visit to G\"ottingen, Einstein was clearly delighted by the response he received:  ``to my great joy, I succeeded in convincing
Hilbert and Klein completely.''\footnote{Einstein to
W.J. de Haas, undated,
probably August 1915, \cite[162]{Einstein1998a}.} As for Hilbert's reaction to Einstein's visit, this encounter inspired him to consider whether general relativity 
might provide a fruitful framework for combining Einstein's gravitational theory with Gustav Mie's electromagnetic theory
of matter.

By the fall of 1915, however, Einstein was no longer expressing the kind of 
 self-satisfaction he felt immediately after
delivering his G\"ottingen lectures.
On 7 November, he wrote
Hilbert: ``I realized about four weeks ago
that my methods of proof used until then were deceptive'' \cite[191]{Einstein1998a}.
Thus began a flurry of exchanges in which Einstein and
Hilbert corresponded directly with
one another as well as through Arnold Sommerfeld \cite{Rowe2001}.
On November 20, Hilbert presented
 the first of his two communications 
 to the
 G\"ottingen Scientific Society.
 Five days later, Einstein submitted \cite{Ein-1}, the last of his four
 notes on general relativity to the Berlin
 Academy.
 Abandoning the basic
assumptions of his theory, he reaffirmed the centrality of
general covariance while seeking a corresponding set of
field equations for gravitation by making use of the Ricci tensor.\footnote{On Einstein's struggles from this period, see
\cite{Stach}, \cite{Janssen2014}, \cite{Janssen/Renn2007}, and \cite{Janssen/Renn2015}.}

 The note \cite{Ein-1} contains the fundamental equations:
\begin{equation}\label{eq:einst}
 R_{\mu\nu}= -\kappa(T_{\mu\nu}-\frac{1}{2}g_{\mu\nu}T),
\end{equation}
 where   $g_{\mu\nu}$ and $R_{\mu\nu}$   are the metric and Ricci
  tensors, respectively. Einstein's argument for these equations was
	highly heuristic in nature, but he had already shown how, by using them 
	in a simplified form, they could be used to 
calculate the displacement in Mercury's perihelion, a major breakthrough for precision 
measurements in solar astronomy. Hilbert, on the other hand, was able to derive gravitational field equations from a variational principle, an important mathematical achievement. Much as he had done in his other physical research, Hilbert hoped that by exploiting  axiomatic and variational methods he would be able to place relativistic field theory on a firm footing.\footnote{While some authors have portrayed the events of November 1915 as a race to arrive at the equations (\ref{eq:einst}), recent research has made clear that this was indeed a major concern for Einstein, but much less so for Hilbert; see  \cite{Sauer1999} and 
\cite[9--17]{Sauer/Majer2009}.} 

Initially, Emmy Noether worked closely with Hilbert, but she also assisted Felix Klein in preparing his lectures on the
development of  mathematics  during the nineteenth century  \cite{Kl-5}.
Starting in the summer of 1916, Klein broke off these lectures in order to begin a 3-semester course  
 on the mathematical foundations of
 relativity theory (published posthumously in \cite{Kl-6}). Much of what he presented during the first two semesters centered on the background to special relativity, including Maxwell's theory, but also the classical theory of algebraic and differential invariants. By the third semester, though, he entered the mathematical terrain of general relativity: Riemannian geometry and Ricci's absolute differential calculus.
Klein's interests diverged rather strikingly from those of Minkowski and Hilbert, both of whom hoped to break new ground in electrodynamics. Unlike them, he was exclusively interested in the mathematical underpinnings of the new physics.
Once Einstein pointed the way to a gravitational theory based on generalizing Minkowski space to a Riemannian manifold, Klein began  to explore the purely mathematical foundations underlying Einstein's new Ansatz.\footnote{One year after
Minkowski's premature death in 1909, Klein took up the connection between
Minkowski's spacetime geometry, based on the invariance properties of the
Lorentz group, and the ideas in his ``Erlangen Program'' \cite{Kl-3}.
Klein's ``Erlangen Program'' \cite{Kl-4}
was republished many times, e.g. in
\cite[460--497]{Kl-GMA}.} 
By the end of 1917, Klein sent Einstein a copy of the
{\it Ausarbeitungen} of his lectures on the
mathematical foundations of relativity. The latter's
response was not very flattering:
``it seems to me that you highly overrate the value of formal points of view. These
may be valuable when an {\it already found} truth needs to be
formulated in a final form, but they
fail almost always as heuristic aids'' \cite[569]{Einstein1998a}.

 Compared with Hilbert's research program, Klein's 
 agenda was rather 
 modest. Indeed,  
Hilbert was pursuing the far more ambitious goal of trying to find a connection between gravity and electromagnetism. His guiding ideas regarding the latter came from Gustav Mie's theory of matter.\footnote{Mie's approach to field physics also exerted a strong influence on Hermann Weyl up until around 1920. See Weyl's remarks in \cite[1952: 211]{Weyl-2} and the note he later added on p. 216 after he became disillusioned with this program.}
Hilbert was especially attracted to Max Born's presentation of Mie's theory in \cite{Born1914}\footnote{For discussions of this paper, see
\cite[309--315]{Corry2004} and \cite{Smeenk/Martin2007}.} because of its 
mathematical elegance and reliance on variational methods. Variational principles had a longstanding place in classical mechanics, particularly due to the influential work of J. L. Lagrange, but their use in electrodynamics and field physics brought about numerous challenges.
In the context of Mie's theory, Born showed how to derive its fundamental equations from a variational principle by varying the field variables
rather than varying the coordinates for space and time. 
 
Emmy Noether presumably had little knowledge of variational methods when she joined Hilbert's research group in 1915. What she knew very well, however, were related methods for using formal differential operators to generate algebraic and differential invariants.\footnote{For an introduction to this field, see \cite{Olver}.} In November 1915 she wrote to her friend and former Erlangen mentor, Ernst Fischer, to tell him about her work in G\"ottingen. Fischer had studied in Vienna under Franz Mertens, a leading expert on invariant theory whose work had influenced the young
 David Hilbert.\footnote{Mertens influence on Hilbert is recounted in \cite[163--164]{Rowe2018}.} From Noether he now learned that 
 Hilbert had created a buzz of excitement about invariant theory, so that even the physicist Gustav Hertz was studying
the classical literature \cite[30--31]{Dick}. She herself had learned these older methods in Erlangen from Paul Gordan, who supervised her dissertation, a tedious study of the  invariants and covariants associated with a ternary biquadratic form. 
 Hertz was learning them from 
her {\it Doktorvater's} old lectures, edited by Georg Kerschensteiner in \cite{Kersch}. Noether knew that Hilbert was pushing his team on with hopes for a breakthrough in physics, but she freely admitted that none of them had any idea what good their calculations might be \cite[30--31]{Dick}.

No doubt Hilbert thought about this along lines first explored by Gustav Mie in his search for a 
suitable ``world function'' $\Phi$ that would lead to an electromagnetic theory of matter
\cite{Mie1912}. Mie assumed that such a function $\Phi$  would have to be Lorentz covariant, thus compatible with the special theory of relativity, and furthermore that it should depend on the field variables alone. As Max Born pointed out, the latter assumption represented an important deviation from classical electron theory, in which the space and time coordinates enter the Lagrangian formalism. ``In Mie's theory,'' he writes ``the forces that hold atoms and electrons together should arise naturally from the formulation of $\Phi$, whereas in the classical theory of electrons the forces have to be specifically added'' \cite[753]{Born1914}. As for the demand that $\Phi$  be Lorentz covariant, Mie showed that this meant it had to be a function of just four invariant quantities.\footnote{Several years later, it was discovered that Mie and his contemporaries had overlooked a fifth invariant; see \cite[627, footnote 9]{Smeenk/Martin2007}.} 

This same feature applied to Hilbert's world function $H$, which was invariant under general coordinate transformations. In his lecture course on foundations of physics from 1916/17, Hilbert emphasized the importance of restricting the possibilities for 
the Lagrangian $H$ \cite[287--290]{Sauer/Majer2009}. He took this to be of the form $H=K+L$, where 
$K$ is the Riemannian curvature scalar and the electromagnetic Lagrangian $L$ depends on the metric tensor $g_{\mu\nu}$, but not on its derivatives. Hilbert noted that the $g_{\mu\nu}$ had to be present in $L$, as otherwise one could not construct any invariants from the electromagentic potentials alone. By the same token: ``this assumption leads to a truly {\it powerful simplification},'' since it means that $L$ has to be a function of just four known invariants \cite[287]{Sauer/Majer2009}. Since the gravitational part of $H$ was given by $K$, this meant that Hilbert's program rested on finding the requisite properties of these invariants in order to construct $L$.\footnote{In this course from 1916/17, Hilbert made the additional assumption that the derivatives $q_{hk}$  only enter $L$ quadratically, from which he deduced that $L$ takes the form $L=aQ+a_1Q_1+a_2Q_2+ f(q)$, where $Q,\,Q_1,\,Q_2, \,q$ are the four known invariants underlying his theory.}
His initial enthusiasm for these ideas did not last long, however, and by 1917 Hilbert's ambitions for a unified field theory
of ``everything'' passed over to a ready acceptance of Einstein's position, namely that general relativity had no immediate relevance for microphysics \cite{Renn/Stachel2007}.

 Emmy Noether presumably gained some understanding of what Hilbert hoped to achieve for physics by drawing on advances in invariant theory. Yet if so, she surely never thought of her own work as motivated by Hilbert's physical program. 
In fact, she was already pursuing a program for invariant theory that was inspired by her collaboration with Ernst Fischer.\footnote{For an idea of the scope of her research program in invariant theory, see \cite{Noether1923}.}
On 22 August 1917, she wrote him to announce that she had finally solved a problem that had occupied her attention since spring, namely the extension of a theorem proved by E.B. Christoffel and G. Ricci for quadratic differential forms to forms with any finite number of variables \cite[33]{Dick}.  On 15 January 1918, Noether presented a lecture on her ``Reduction Theorem'' 
 at a meeting of the G\"ottingen Mathematical Society, and ten days later Felix Klein submitted her paper \cite{Noether1918a} for publication. Drawing on methods in the calculus of variations introduced by Lagrange, Riemann, and Lipschitz, she shows how problems involving systems of differential invariants can be reduced to classical invariant theory, i.e. invariants of the projective group. Her treatment of Lagrangian derivatives as formal invariants reveals that this paper is closely related to the far more famous \cite{Noether1918b}.

\section{On Conservation Laws in General Relativity}

Whereas Hilbert hoped to use Einstein's gravitational theory as a framework for a new unified field theory, Noether remained what she had always been: a pure mathematician. Her work thus 
 aimed to clarify the mathematical underpinnings of general relativity, an effort strongly promoted by Felix Klein, who took up this challenge around the time that Hilbert's interests were turning back to the foundations of mathematics \cite[22]{Sauer/Majer2009}.
 In March of 1917, Klein
initiated a correspondence with Einstein
that sheds considerable light on
how both
thought about relativity theory and the
general relationship between
mathematical and physical reasoning.
Their letters mainly reflect the
three topics which were then uppermost in Klein's mind:
1) the conceptual links between relativity theory and
his ``Erlangen Program''; 2) the cosmological models proposed by Einstein and Willem de Sitter, in particular as these related to 
non-Euclidean geometries;
and 3) the role of conservation
laws in classical and relativistic physics. Only this last topic will be discussed here, but the others are suggestive of the broader range of issues central to the reception of general relativity in G\"ottingen.\footnote{For a discussion of topic 2), see \cite[279--299]{Rowe2018}.}  

 Beginning in March 1918, the correspondence between Klein and Einstein
 intensified
markedly
following the appearance of \cite{Kl-1}.
 This paper arose from a presentation Klein made on 22 January 1918 to members of the G\"ottingen
 Mathematical Society, a talk that elicited a reaction
 from Hilbert one week later. The conclusions drawn from these two
 sessions were later summarized in the journal of the German
 Mathematical Society: ``The `conservation laws' valid for continua
 in classical mechanics (the impulse-energy theorems) are already
 contained in the field equations in
 Einstein's newly inaugurated theory; they thereby lose their
 independent
 significance.''\footnote{{\it Jahresbericht der
 Deutschen Mathematiker--Vereinigung}, 27 (1918),
 (``Mitteilungen und Nachrichten''), p. 28.} 

Klein and Hilbert afterward agreed to publish their respective viewpoints in the {\it G\"ottinger Nachrichten}  
as an epistolary exchange.\footnote{Klein had already presented a preliminary version of \cite{Kl-1} at a meeting of the scientific society on 25 January.} Considering that they both lived only a short distance from one another on the Wilhelm Weber Strasse, one might wonder why they chose to publish the gist of their discussions as an exchange of correspondence. In any event, the views they set forth harmonized and were surely meant to be seen as representing 
 the consensus opinion on these matters in G\"ottingen.
In \cite{Kl-1}, Klein underscored that Hilbert's invariant energy equation should not be viewed as a conservation law in the sense of classical mechanics. The latter could only be derived by invoking physical properties of matter, whereas Hilbert's equation followed directly from the gravitational field equations by means of purely formal considerations. Klein further remarked that Emmy Noether had already noticed this and had worked out all the details in a manuscript, a text she had shown him.
  ``You know,'' he wrote, ``that Miss Noether advises me continually regarding my work, and that, in fact, it is only thanks to her that I have understood these questions'' \cite[559]{Kl-1}.

Hilbert was certainly very well aware of this, and he responded as follows: 

\begin{quote}

I fully agree with the substance of your statements on the energy theorems. Emmy Noether, on whom I have called for assistance more than a year ago to clarify this type of analytical question concerning my energy theorem, found at that time that the energy components that I had proposed -- as well as those of Einstein -- could be formally transformed, using the Lagrangian differential equations . . . of my first note, into expressions whose divergence vanishes identically . . .. \cite[560--561]{Kl-1}.
\end{quote}

As it happens, a Swiss student named Rudolf Jakob Humm attended Klein's lecture and was impressed by what he heard. 
Humm was studying relativity under Hilbert, but he had also spent a semester in Berlin attending Einstein's lecture course. Since he was particularly interested in energy conservation, Humm surely read \cite{Kl-1}, which would have made him aware of Noether's manuscript had he not known about it already before. In any event, Humm must have approached her at some point to ask if he could copy part of this text.\footnote{This copy, written in Humm's hand, is just one of the many documents in his posthumous papers relating to his interest in general relativity during the war years (Nachlass Rudolf Jakob Humm, Zentralbibliothek Z\"urich).} 

Humm  grew up in Modena and later
completed his secondary education at the Kantonsschule in
Aarau. This was the same
 institution Einstein had attended for one year
before he entered the Polytechnicum in Z\"urich, a
circumstance that possibly inspired Humm to study relativity
in Germany. He first studied mathematics in Munich in 1915, before moving on to
G\"ottingen. By the winter semester of 1916/17 he was
thoroughly steeped in theoretical physics. 

During wartime, university enrollments plummeted, so Humm was moving in a 
small world in which people saw one another nearly every day. His contacts with 
 Emmy Noether were apparently
rather fleeting, whereas he regularly interacted with several fellow natives of Switzerland, including  
the physicist Paul Scherrer and his wife, Paul Finsler, and
Richard B\"ar. Humm also socialized with Vsevolod
Frederiks, one of several Russians studying physics and
mathematics in G\"ottingen \cite[114--115]{Rowe2004}, and he befriended 
 Willy Windau, a blind mathematician who went on
to take his doctorate under Hilbert in 1920. Another sometime companion
was the astronomer Walter Baade, who took his
degree in 1919.

One evening in
April 1917, he and Baade met for drinks at
the Hotel National. Humm was
somewhat despondent on this occasion, in part because of
the meager course offerings for the coming semester. He
had been following Hilbert's relativity course with enthusiasm,
but for the summer, the master would be teaching
only a four-hour course on set theory. Over the course of that evening, Baade
convinced him to leave G\"ottingen for Berlin, where 
Einstein had already begun
teaching a course on relativity. A few days later, Humm was already settled in and looking forward
to Einstein's course, which was held on Thursdays from 2 to 4. He also made plans to
attend the physics colloquium, which was run by Heinrich Rubens.

Humm had missed the first two lectures, so he had some
questions after hearing the third. Evidently, Einstein offered
to meet with him the following Saturday, an encounter that
led to a series of remarks by the physicist that Humm tried
to reconstruct in his diary. Einstein had recently read Hilbert's
second note on the foundations of physics \cite{Hilbert1917}. There, Hilbert had introduced a special coordinate
system in order to preserve causal relations in general relativity,
but Einstein thought this was inadmissible because
it could lead to worldlines that converge, thereby yielding
space-time singularities. He had already mentioned this
criticism two weeks earlier in a letter to Felix Klein.\footnote{Einstein to Klein, 24
April 1917, \cite[426]{Einstein1998a}.}

This was only one of several conversations Humm had
with Einstein during his three-month stay in Berlin.
Alongside Einstein's course, he also attended Max Planck's
lectures on quantum theory as well as Rubens's weekly
colloquium, which met on Wednesdays. He found this all
quite stimulating, but he also missed the conveniences of
G\"ottingen's Lesezimmer. In Berlin one had to order books
from the library, so there was no opportunity to browse
open shelves to pick out the volumes one might want to
read. Rubens had asked Humm to speak in the colloquium
in early August, but this plan was aborted after Einstein fell
ill in mid-July. His assistant, Jakob Grommer, then took
over the course, while Einstein left for Switzerland to
recover from an intestinal ailment. For Humm, this sudden
turn of events meant that he had little incentive to stay in
Berlin any longer. So he canceled his colloquium lecture
and soon thereafter returned to G\"ottingen.

Humm was surely well-versed when it came to the various proposals for dealing with energy conservation in general relativity. He had attended Hilbert's year-long course on ``Die Grundlagen der Physik'', in which this topic received renewed attention \cite[304--306]{Sauer/Majer2009}. When he arrived back 
in G\"ottingen for the winter semester 1917/18, Hilbert appointed him to prepare the official {\it Ausarbeitung} for his lecture course on electron theory.
At the end of Einstein's final lecture, Humm was keen to learn his opinion about 
one of the most controversial parts of his theory, namely Einstein's pseudo-tensor for representing gravitational energy. Throughout 1917 and 1918, Einstein argued against much skeptical opinion that the expression for gravitational energy could not be a general tensor; on the contrary, it needed to vary with the coordinate frame. Humm recorded this response in his notes from Einstein's lecture:

\begin{quote}

I asked Einstein if it would be possible to generalize the conservation equation
$$\frac{\partial{({\frak T}^{\sigma}_{\mu}+{\frak t}^{\sigma}_{\mu})}}{\partial{x_{\sigma}}}=0$$
 so that it would contain only real tensors. He thought not: one does not shy from writing            
$$\frac{\partial{(T+U)}}{\partial{t}}=0.$$
in classical mechanics, where $U$ in an invariant under Galilean transformations, but $T$ is not. So it not so terrible to have the general tensor ${\frak T}_{\mu}^{\sigma}$       next to the special  ${\frak t}_{\mu}^{\sigma}$     . If one considers an accelerative field, then there will be a   ${\frak t}_{\mu}^{\sigma}$    , even though the field can be transformed away. In the end, one can operate with any arbitrary concept, and it cannot be said that they have to be tensor quantities; the [Christoffel symbols] are also not tensors, but one operates with them. The  ${\frak t}_{\mu}^{\sigma}$     are the quantities that deliver the most. (Nachlass Rudolf Jakob Humm, Zentralbibliothek Z\"urich)

\end{quote}

Humm was strongly drawn to Einstein's highly conceptual
way of thinking about fundamental physical
problems, an approach he contrasted with Hilbert's purely
mathematical approach. Energy conservation and the
equations of motion in general relativity would thenceforth
become his principal research agenda.

Meanwhile, Humm continued to stay in contact with
Einstein, who submitted two of his papers for publication
in {\it Annalen der Physik}. The first of these was  \cite{Humm1918},
written in May 1918, just two months before Emmy Noether
presented her paper \cite{Noether1918b}. In it, Humm takes
considerable care to explain how one can apply different
variational methods to obtain results adapted to a particular
physical setting. Among his findings, he could show that
Einstein's differential equations for conservation of energy
were derivable from the equations of motion, i.e., the
assumption that a test particle moves along a geodesic in
curved space-time. Humm submitted 
\cite{Humm1919}, his second paper, just one month after Noether completed hers;
again, one finds striking parallels between them. This
second contribution aimed to show that Einstein's energy
equations could be seen as equivalent to equations of
motion, an argument based on certain analogies with
Lagrangian mechanics.

Humm's transcription of  Noether's results from 1916 will be discussed below. First, however, it will be necessary to consider how Hilbert derived his invariant energy vector in \cite{Hilbert1915}, after which I will briefly describe Einstein's handling of energy conservation in general relativity in \cite{Ein-6}.

\section{Hilbert's Approach to Energy in \cite{Hilbert1915} }

When Einstein delivered his six Wolfskehl lectures in G\"ottingen in the late spring of 1915, he was advancing a version of a gravitational theory that he had earlier worked out with the help of the mathematician Marcel Grossmann \cite{E/G1913}. At this time, he was convinced that  if such  a theory were based entirely on the principle of general covariance, then it would necessarily be undetermined. For this reason, he and Grossmann  reached the conclusion that the gravitational field equations could not be generally covariant. Instead, their equations were covariant only with respect to a more restricted group that included the linear transformations.  In G\"ottingen Einstein quite possibly spoke  about the possibility of using energy conservation in order to bring about this restriction, one of several problems he had yet to solve. 

In any event, Hilbert's initial attempt to subsume Mie's theory within the context of general relativity followed  
Einstein's then current belief that the field equations themselves could not be 
generally covariant. Moreover, to avoid this problem he struck on the idea of utilizing 
 energy conservation to restrict the system of allowable coordinates, a method designed to preserve causal relations. This initial foray into general relativity, however, never found its way into print. Hilbert's contemporaries were therefore unaware that his original approach to energy conservation differed fundamentally from the one that appeared in his  published note, \cite{Hilbert1915}. The discrepancy was only discovered in the late 1990s when  historians discovered that, although this note still bore the original date of submission (20 November 1915), 
Hilbert had  heavily revised it after receiving the  page proofs  in December 1915 \cite{Cor}.\footnote{The extant page proofs are incomplete, however. What they likely once contained has been discussed in \cite{Sauer2005} and in \cite{Renn/Stachel2007}.}

In these page proofs, published as \cite{Hilbert2009},  Hilbert introduces a linear invariant as the ``energy form,''  from which  he derives four coordinate conditions from a divergence equation. He then introduces this coordinate system as an ``axiom for space and time,'' thereby  obtaining a total of 14 differential equations for the 14 field variables required for combining Einstein's and Mie's theories. He adopted this strategy in order to circumvent the problem he foresaw if the theory only admitted 10 equations, thereby leaving four degrees of freedom for the motion of a physical system. 
Hilbert dropped all  this, however, in the published version of \cite{Hilbert1915}, where his modified energy law is fully covariant. 
Nevertheless, as described in 
\cite{Br-Ryck2018}, he continued to struggle with the problem of reconciling general covariance with causality, which returns to the fore in \cite{Hilbert1917}, the sequel to his first note on ``Die Grundlagen der Physik''. As John Stachel notes in 
\cite{Stachel1992}, this paper was the first attempt to deal with the Cauchy problem in general relativity. In it, Hilbert commented:

\begin{quote}
As far as the causality principle is concerned, if the physical quantities and their time derivatives are known in the present in any given coordinate system, then a statement will only have physical meaning if it is invariant with respect to those transformations for which the coordinates used are precisely those for which the known present values remain invariant. I claim that all assertions of this kind are uniquely determined for the future as well, i.e., that the causality principle is valid in the following formulation:
From knowledge of the fourteen potentials \dots \,\,
 in the present all statements about them in the future follow necessarily and  uniquely insofar as they have physical meaning.  
\cite[61]{Hilbert1917}
\end{quote}

Hilbert's theory in \cite{Hilbert1915} was based on 
 two axioms that concern the properties
of a ``world
function'' $$H(g_{\mu\nu}, g_{\mu\nu, l}, g_{\mu\nu, lk}, q_s, q_{s,l}).$$ This
$H$ is taken to be 
a scalar-valued function that does not depend explicitly on the spacetime coordinates $w_s$   but rather on
 the ten components of the symmetric metric tensor 
 $g_{\mu\nu}$ and its first and second derivatives as
 well as four electromagnetic potentials $q_s$ and their first
 derivatives. Hilbert notes that $H$ could just as well be defined by means of the contravariant arguments $g^{\mu\nu}, \, g^{\mu\nu}_l,
\, g^{\mu\nu}_{lk}$, which he adopts afterward.
Axiom I then 
  asserts that
under infinitesimal variations of the field functions $g^{\mu\nu}
 \rightarrow g^{\mu\nu} +\delta g^{\mu\nu}$ and $q_s \rightarrow q_s
 + \delta q_s$,
$$\delta\int H\sqrt g d\omega =0,$$
 where $g= | g^{\mu\nu}|$ and $ d\omega  = dw_1dw_2dw_3dw_4.$
 This variational principle is understood to apply throughout
 a finite region of space-time.
 Axiom II then simply states that this world function $H$ is taken to be
 invariant under general coordinate
 transformations.

By virtue of Axiom I, Hilbert obtained ten Lagrangian differential equations for the
ten gravitational potentials $g^{\mu\nu}$:
\begin{equation}\label{eq:grav-Lagrange}
\frac{\partial \sqrt g H}{\partial g^{\mu\nu}} - \sum_k
\frac{\partial}{\partial w_k}\frac{\partial \sqrt g H}
{\partial g_k^{\mu\nu}} + \sum_{k,l}\frac{\partial^2}
{\partial w_k \partial w_l} \frac{\partial \sqrt g H}
{\partial g_{kl}^{\mu\nu}}=0.
\end{equation}
Similarly, for the four electrodynamic potentials $q_s$, he derived the four equations:
\begin{equation}\label{eq:ed-Lagrange}
\frac{\partial \sqrt g H}{\partial q_h} - \sum_k  \frac{\partial}
{\partial w_k}\frac{\partial \sqrt g H}{\partial q_{hk}} =0.
\end{equation}
Hilbert called the first set of equations the fundamental equations
of gravitation and the second the fundamental equations of
electrodynamics,  
abbreviating these to read:
\begin{equation}\label{eq:Lagrange-derivs}
[\sqrt g H]_{\mu\nu} =0, \quad [\sqrt g H]_h =0.
\end{equation}

In the course of developing his theory, Hilbert focused on the special case where the Lagrangian $H$ takes the form $H= K+L$.
Here $K$ is the curvature scalar obtained by contracting the Ricci tensor $K_{\mu\nu}$, i.e., 
$K= \sum_{\mu\nu} g^{\mu\nu}K_{\mu\nu}$.
 He placed no special conditions on the Lagrangian $L$, but noted that  it contained no derivatives of the metric tensor, so that
 $H = K + L(g^{\mu\nu}, q_s, q_{s,l})$. Utilizing a general theorem for constructing differential invariants from a given invariant, Hilbert showed how $L$ led to a differential equation that served as a generalized Maxwell equation, as in Gustav Mie's electromagnetic theory of matter.

Hilbert's central claim concerned four independent linear combinations satisfied by the $[\sqrt g L]_h$ and their derivatives. He showed toward the end of \cite{Hilbert1915} how these can be derived from the fundamental equations of gravitation. His argument depended on first identifying the electromagnetic part of his energy vector $e^l$ with Mie's expression for energy. Hilbert's more general expression passed over to Mie's when the metric tensor $g_{\mu\nu}$ took on values for a flat spacetime. He then showed that this same expression could be linked to the gravitational field equations, which for $H=K+L$ take the form:
\begin{equation}\label{eq:Hilbert-grav-eqns}
[\sqrt g K]_{\mu\nu} +  \frac{\partial \sqrt g L}{\partial g^{\mu\nu}} =0.\footnote{Much has written about how Hilbert came to recognize the Lagrangian derivative $[\sqrt g K]_{\mu\nu}$ as being identical with the Einstein tensor $\sqrt g (K_{\mu\nu}-\frac{1}{2}g_{\mu\nu}K)$, but it should not be overlooked that this claim  plays no role whatsoever in the arguments presented in \cite{Hilbert1915}. Hilbert surely felt it important to establish this linkage with Einstein's theory of gravitation, but the results he set forth did not make use of the specific form of Einstein's field equations.}
\end{equation}
The electromagnetic part of $e^l$ could then be written:
\begin{equation}\label{eq:Hilbert-EM-Energie}
\frac{-2}{\sqrt g}\sum_{\mu, s}\frac{\partial \sqrt g L}{\partial g^{\mu s}}g^{\mu l}p^s.
\end{equation}
After carrying out a number of quite complicated transformations and making use of (\ref{eq:Hilbert-grav-eqns}), Hilbert obtained four identities involving the Lagrangian expressions
$ [\sqrt g L]_m$ and their first derivatives:
\begin{equation}\label{eq:Lagrange-identities}
\sum_{m}(M_{\mu\nu}   [\sqrt g L]_{m} + q_v \frac{\partial  [\sqrt g L]_{m}}{\partial w_m}) =0,\footnote{His derivation of this equation is found on \cite[405--406]{Hilbert1915}.}
\end{equation}
where $M_{\mu\nu}  = q_{\mu\nu}-q_{\nu\mu}.$

 Somewhat misleadingly, however, he related this  specific result to a much more general one, announced but not proved at the outset. This was his Theorem I, which generalized the situation described by axioms I and II to any invariant $J$ in $n$ variables and their derivatives. From this invariant variational framework one can then derive $n$ Lagrangian differential equations, from which $n-4$ of these lead to four identities satisfied by the other four and their total derivatives. So stated, this theorem asserts that 
four of the fourteen equations $[\sqrt g H]_{\mu\nu} =0 \,\, [\sqrt g H]_h =0$
can be deduced directly from the other ten. Hilbert seized on this result to make a strong physical claim:
\begin{quote}

\dots on account of that theorem
we can immediately make the assertion, {\it that in the
sense indicated the electrodynamic phenomena are the effects of gravitation.}
 In recognizing this, I discern the simple and very surprising solution of
 the problem of Riemann, who was the first to search for a theoretical
 connection between gravitation and light. \cite[397--398]{Hilbert1915}\footnote{Hilbert was alluding here to Riemann's
posthumously published ``Gravitation und Licht,''
in \cite[496]{Riem}.}

\end{quote}
Hilbert would later drop this passage in \cite{Hilbert1924}, a new version of his two notes, although the theorem he stated was correct and certainly important. 

A more immediately controversial and confusing aspect, however, concerned Hilbert's handling of energy conservation. In the first part of his paper, he constructed a complicated invariant $e^l$, his energy vector, from which he proved that its divergence vanished; this was his invariant energy theorem:

\begin{equation}\label{eq:Hilbert-energy}
\sum_l  \frac {\partial \sqrt g e^l}
{{\partial w_l}  } =0.
\end{equation}
The energy vector $e^l$  is defined by starting with an arbitrary vector $p^l$ and then 
building four other vectors by means of differential invariants. 
The resulting construction takes this form: 
\begin{equation}\label{eq:Hilbert-vector}
e^l= H\, p^l-a^l-b^l-c^l-d^l.
\end{equation}
Each of these five terms is an invariant, but only the first depends on both the gravitational and electromagnetic potentials. The vectors $a^l, \, b^l$ contain expressions without the $q_s$, whereas the $c^l, \, d^l$ are independent of the $g^{\mu\nu}$. Hilbert 
emphasized that his energy equation holds for any $H$ satisfying the first two axioms, even though the construction of $e^l$ clearly reveals that he had the special case $H=K+L$ in mind. Thus, while he formulates Theorem II for a general invariant $J$ of the type $H$, Hilbert decomposes the operator $P$ that acts as the first polar:

\begin{equation}\label{eq:polar}
\sum_l  \frac {\partial J}
{\partial w_s} p^s  =P(J)\footnote{This equation vexed Einstein, who wrote to Hilbert on 25 May 1916 (see below). Hilbert noted in his reply that the coefficients in the power series expansion arising from a displacement of the variables in the invariant $J$ will themselves be invariant, and furthermore that the first order terms are those given by $P=P_g+P_q$. }
\end{equation}
by writing $P=P_g+P_q$ in order to separate the gravitational and electromagnetic terms, where
$$P_g= \sum_{\mu,\nu,l,k } (p^{\mu\nu}\frac {\partial }{\partial g^{\mu\nu}}+p^{\mu\nu}_l\frac {\partial }{\partial g^{\mu\nu}_l}+p^{\mu\nu}_{lk}\frac {\partial }{\partial g^{\mu\nu}_{lk}})$$ and
$$P_q= \sum_{l,k } (p_l\frac {\partial }{\partial q_l}+p_{lk}\frac {\partial }{\partial q_{lk}} ),$$
where the lower suffixes in $p$ denote coordinate derivatives. 

Hilbert then applies the first operator to polarize the expression $\sqrt g H$:
\begin{equation}\label{eq:polar-g}
P_g(\sqrt g H) = \sum_{\mu,\nu,l,k } (p^{\mu\nu}\frac {\partial \sqrt g H}{\partial g^{\mu\nu}}+p^{\mu\nu}_l\frac {\partial \sqrt g H}{\partial g^{\mu\nu}_l}+p^{\mu\nu}_{lk}\frac {\partial \sqrt g H}{\partial g^{\mu\nu}_{lk}}).
\end{equation}
Using formal properties of tensors, Hilbert introduces the two vectors $a^l, \, b^l$ and shows that they satisfy the equation

$$P_g(\sqrt g H) - \sum_{l} \frac {\partial \sqrt g (a^l + b^l)}{\partial w_l} = \sum_{\mu ,\nu} [ \sqrt g H]_{\mu\nu} p^{\mu\nu} .$$
In constructing the vector $a^l$, he first notes that the coefficient of $p^{\mu\nu}_{lk}$ in the expression for (\ref{eq:polar-g}), namely $\frac {\partial \sqrt g H}{\partial g^{\mu\nu}_{lk}}$,  is a mixed fourth-order tensor, which enables him to produce $a^l$ by multiplying this tensor with another of the third rank. Emmy Noether would point out in 1916 that the second derivatives of the metric tensor that appear in the definition of $a^l$ cannot be eliminated by means of the field equations; she took this as an indication that Hilbert's energy vector was not analogous to a first integral in classical mechanics. 

By an analogous argument using the second operator, Hilbert obtains the vector $c^l$ which satifies the equation:
$$P_q(\sqrt g H) - \sum_{l} \frac {\partial \sqrt g (c^l)}{\partial w_l} = \sum_{k} [ \sqrt g H]_{\mu\nu} p_k .$$
Adding these two equations and applying the fundamental field equations (\ref{eq:Lagrange-derivs}), it follows that
$$P(\sqrt g H) = \sum_{l} \frac {\partial \sqrt g (a^l + b^l +c^l)}{\partial w_l}.$$
Hilbert now applies the identity (\ref{eq:polar}) to this equation to obtain
$$P(\sqrt g H) = \sum_{s} \frac {\partial \sqrt g Hp^s}{\partial w_s},$$ which leads immediately to the divergence equation:

$$\frac {\partial }{\partial w_l}\sqrt g (Hp^l-a^l-b^l-c^l)=0.$$ To complete the construction of the energy vector (\ref{eq:Hilbert-vector}), Hilbert defined $d^l$ by making use of the skew symmetric tensor
 $\frac {\partial H}{\partial q_{lk}}-\frac{\partial H}{\partial q_{kl}}$. Since this $d^l$ has vanishing divergence, it follows immediately that the vector $e^l$ does as well, which completes his proof of (\ref{eq:Hilbert-energy}).

Hilbert wrote at the outset of this derivation that this was a fundamental result for his theory and that it followed from his two axioms alone, though he of course also made use of Theorem II. His readers must have been quite mystified, however, by the fact that he derived another result, Theorem III, before taking up his energy theorem.\footnote{Hilbert used Theorem III to show how electromagnetic energy (\ref{eq:Hilbert-EM-Energie}), expressed in terms of the derivatives of $L$ with respect to the gravitational potentials $g^{\mu\nu}$, leads by virtue of the gravitational equations (\ref{eq:Hilbert-grav-eqns}) to the identities (\ref{eq:Lagrange-identities}).} This theorem plays no role in Hilbert's treatment of energy conservation but, as we shall see below, it forms the starting point for Emmy Noether's analysis of Hilbert's energy vector.

\section{Einstein's Approach to Energy Conservation}

As is well known, Einstein originally  introduced gravitational effects into his special theory of relativity (SR) by means of the equivalence principle. Once he accepted Minkowski's approach to SR, he eventually found a way to adapt the equivalence principle to it.  In SR, force-free motion in an inertial frame of reference takes place along a straight-line path with constant velocity.
Viewed from a non-inertial frame, on the other hand, this path of motion will be a geodesic curve in a flat spacetime
 \begin{equation}\label{eq:geodesic}
\frac{d^2x_{\tau}}{ds^2}= \Gamma^{\tau}_{\mu\nu}\frac{dx_{\mu}}{ds}\frac{dx_{\nu}}{ds},
\end{equation}
since this equation is independent of the coordinate system. Einstein made the plausible assumption that this geodesic motion   also holds in the non-flat case, i.e. in a spacetime region for which it is impossible to find a coordinate system that leads to the Minkowski metric in SR.\footnote{A number of investigators, including Hermann Weyl, afterward showed how the geodesic equation for motion could be deduced from the field equations (see \cite{Havas1989}). Einstein, however, was reluctant to follow this lead for reasons discussed in \cite{Kennefick2005} and \cite{Lehmkuhl2017}.} This geometrical assumption served as the starting point for his gravitational theory; afterward it stood as a sturdy bridge that joined the special and general theories of relativity.

Einstein's classic paper \cite{Ein-6} was published as a separate brochure that came out just before Einstein began an interesting 
 correspondence with Hilbert, which will be discussed below.
In \cite{Ein-6} one encounters a number of arguments leading to different formulations of the gravitational field equations. 
 From the outset, Einstein posed the unimodular coordinate condition $\sqrt{-g}=1$,\footnote{On Einstein's use of unimodular coordinates, see the discussion in \cite{Janssen/Renn2007}.} which leads to a significant simplification of the field equations (\ref{eq:einst}).
He first considered the matter-free case ($T_{\mu\nu}=0$), 
$R_{\mu\nu}=0$. Here $R_{\mu\nu}$ is the Ricci tensor, which simplifies in unimodular coordinates, so the ten differential field equations can be written 
\begin{equation}\label{eq:modular1}
               \frac{\partial{\Gamma^{\alpha}_{\mu\nu}}}{\partial{x_{\alpha}}}+\Gamma^{\alpha}_{\mu\beta}\Gamma^{\beta}_{\nu\alpha}   =0,
\end{equation}
where the $\Gamma^{\alpha}_{\mu\nu}$ are Christoffel symbols of the second kind (another popular notation is $\Gamma^\sigma_{\mu\lambda} = \left \{ {\sigma \atop
\mu\lambda}\right \} =  -\left \{ {\mu\lambda \atop \sigma
}\right \}$). 

Einstein derives another form of the equations (\ref{eq:modular1}) by using variational methods. Assuming $\sqrt{-g}=1$, he takes the scalar 
\begin{equation}\label{eq:modular-H}
H= \sum_{\alpha\beta\mu\nu}g^{\mu\nu}\Gamma^{\alpha}_{\mu\beta}\Gamma^{\beta}_{\nu\alpha}
\end{equation}
 and 
writes $$\delta \left\{\int H d\tau \right\}=0.$$ Carrying out the variation yields field equations in the form:

\begin{equation}\label{eq:modular2}
\frac{\partial}{\partial{x_{\alpha}}} \left\{\frac{\partial H}{\partial{g^{\mu\nu}_{\alpha}}}\right\} 
-\frac{\partial H}{\partial{g^{\mu\nu}}}=0.
\end{equation}
After a series of intermediate calculations, he obtains:

\begin{equation}\label{eq:modular3}
\sum_{\alpha}\frac {\partial t^{\alpha}_{\sigma}}{\partial x_{\alpha}}=0;\,\, 
-2\chi t^{\alpha}_{\sigma}= \sum_{\mu\nu}\left \{g^{\mu\nu}_{\sigma}\frac{\partial H}{\partial {g^{\mu\nu}_{\alpha}}}\right \}-\delta^{\alpha}_{\sigma}H.
\end{equation}
Einstein noted that although 
$t^{\alpha}_{\sigma}$ is not a general tensor, the equations (\ref{eq:modular3}) are valid whenever $\sqrt{-g}=1$. He interpreted the 
$t^{\alpha}_{\sigma}$ pseudo-tensor as representing the energy components of the gravitational field and (\ref{eq:modular3}) as expressing the equation for conservation of momentum and energy in the vacuum case. For the pseudo-tensor, he derives the equation
\begin{equation}\label{pseudo-tensor}
\chi t^{\alpha}_{\sigma}= \sum_{\mu\nu\lambda}\frac{1}{2}\delta^{\alpha}_{\sigma}g^{\mu\nu}\Gamma^{\lambda}_{\mu\beta}\Gamma^{\beta}_{\nu\lambda}-g^{\mu\nu}\Gamma^{\alpha}_{\mu\beta}\Gamma^{\beta}_{\nu\alpha}.
\end{equation}

Einstein's generalization of (\ref{eq:modular3}) in the presence of a matter tensor $T^{\sigma}_{\mu}$ takes the form
\begin{equation}\label{energy conservation}
\frac{\partial{(T^{\sigma}_{\mu}+t^{\sigma}_{\mu})}}{\partial{x_{\sigma}}}=0.
\end{equation}
He obtains  this by deriving yet another form for the field equations (\ref{eq:modular1}), still assuming the condition 
$\sqrt{-g}=1$:
\begin{equation}\label{eq:modular4}
\sum_{\alpha\beta}\frac{\partial}{\partial{x_{\alpha}}} \left (g^{\sigma\beta}\Gamma^{\alpha}_{\mu\beta} \right ) =-\chi 
\left (t^{\sigma}_{\mu}-\frac{1}{2}\delta^{\sigma}_{\mu}  t   \right ),
\end{equation}
where $t=\sum_{\alpha}t^{\alpha}_{\alpha}$. 

He then modifies equations (\ref{eq:modular4}) by replacing $t^{\sigma}_{\mu}$ with $t^{\sigma}_{\mu}+T^{\sigma}_{\mu}$ to obtain:
\begin{equation}\label{eq:modular5}
\sum_{\alpha\beta}\frac{\partial}{\partial{x_{\alpha}}} \left (g^{\sigma\beta}\Gamma^{\alpha}_{\mu\beta} \right ) =-\chi 
\left [(t^{\sigma}_{\mu}+T^{\sigma}_{\mu})-\frac{1}{2}\delta^{\sigma}_{\mu}(t+T)  \right ].
\end{equation}

By means of (\ref{eq:modular5}) and some intermediate calculations, Einstein derives the differential form for conservation of momentum and energy (\ref{energy conservation}). These results from \cite{Ein-6} clearly differ sharply from the findings in \cite{Hilbert1915} discussed above.  Nevertheless, Emmy Noether was able to show that Hilbert's $e^l$ and Einstein's $t^{\alpha}_{\sigma}$ both possessed a common property which seemed to reflect that the energy laws in general relativity differ from  those in classical mechanics or special relativity.

 Soon after he published \cite{Ein-6} in May 1916, Einstein made a conscientious attempt to understand how Hilbert developed his far more complicated approach to energy-momentum conservation published  in \cite{Hilbert1915}.\footnote{The complications were in part due to the fact that Hilbert decided to alter his definition of energy in the page proofs of his original submission from 20 November 1915. Thus the version   in \cite{Hilbert1915} actually reflects an important shift  in Hilbert's understanding of this aspect of his theory. For details, see Tilman Sauer's commentary in 
\cite{Sauer/Majer2009}, pp. 11--13.}
Einstein struggled to understand the arguments in Hilbert's first note as he prepared to speak about it in Heinrich Ruben's colloquium. Twice he turned to Hilbert for clarifications, writing: ``I admire your method, as far as I have understood it. But at certain points I cannot progress and therefore ask that you assist me with brief instructions'' 
 (Einstein to Hilbert, 25 May 1916, \cite[289]{Einstein1998a}). He was particularly baffled by Hilbert's energy theorem, 
 admitting that he could not comprehend it at all -- not 
even what it asserted.\footnote{Hilbert claimed not only that the energy vector $e^l$ depended solely on the metric tensor and its derivatives,
he also showed that by passing to a flat metric its electromagnetic part turned out to be closely related to a formulation for energy derived from Mie's theory. Einstein was puzzled about this derivation, since the argument seemed to show that not only the divergence of the energy term but this term itself would have to vanish.}
 
Hilbert wrote back just two days later. He easily explained how, via the operation of polarization, an invariant $J$ will lead to a new invariant $P(J)$, its first polar. He then went on to say:

\begin{quote}
My energy law is probably related to yours; I have already assigned this question to Miss Noether. As concerns your objection, however, you must consider that in the boundary case $g^{\mu\nu} = 0,\, 1$ the vectors $a^l,\, b^l$ by no means vanish, as $K$ is linear in the $g^{\mu\nu}_{\sigma\kappa}$ terms and is differentiated with respect to these quantities. For brevity I give you the enclosed paper from Miss Noether. 

\end{quote}

Hilbert's conjecture regarding the relationship between his and Einstein's versions of energy conservation was surely 
 no more than a first guess. Even on the purely formal level, he could hardly assert that his energy vector $e^l$ stood in some obvious relation to Einstein's pseudo-tensor $t^{\alpha}_{\sigma}$.

Einstein was well aware that Noether was working closely with  Hilbert and that the latter had been trying to break the resistance in the faculty to her appointment as a {\it Privatdozent}. Despite strong support by the members of the natural sciences division, however, all such efforts proved impossible during wartime. Only after the fall of the German Reich and the advent of the Weimar Republic did these efforts succeed (see \cite{Tollmien1990}). Einstein responded to Hilbert's letter shortly afterward:

\begin{quote}

Your explanation of equation [(\ref{eq:polar})] in your paper delighted me. Why do you make it so hard for poor mortals by withholding the technique behind your ideas? It surely does not suffice for the thoughtful reader if, although able to verify the correctness of the equations, he cannot have a clear view of the overall plan of the analysis.

\end{quote}

Einstein was far more blunt about this in a letter he wrote to Paul Ehrenfest on May 24:
``Hilbert's description doesn't appeal to me. It is unnecessarily specialized regarding `matter,' is unnecessarily complicated, and not straightforward (= Gauss-like) in set-up (feigning the super-human through camouflaging the methods)'' \cite[288]{Einstein1998a}.
After receiving Hilbert's explanations, he may have felt somewhat more conciliatory. Certainly he made every effort to understand Hilbert's arguments, and could report: ``In your paper everything is understandable to me now except for the energy theorem. Please do not be angry with me that I ask you about this again''   \cite[293]{Einstein1998a}.
After explaining the difficulty he still had, Einstein ended by writing that it
 would suffice if Hilbert asked Emmy Noether to clarify the point that was troubling him. This turned out to be a quite trivial matter, so Hilbert answered Einstein directly. The latter then responded with thanks, adding that ``now  your entire fine analysis is clear to me, also with respect to the heuristics. Our results are in complete agreement''  \cite[295]{Einstein1998a}.

What Einstein meant by this would seem quite obscure. Perhaps he only meant to assure Hilbert that he would no longer be pestering him about these matters. One must assume that Hilbert had just as little interest to enter these waters further, for how else to account for the fact that he failed to publicize Emmy Noether's findings, which clearly stemmed from this correspondence with Einstein? Not until Felix Klein began to take an interest in the status of conservation theorems in GR more than a year later did Noether's name  receive any attention in this connection.

\section{On Noether's Unpublished Manuscript from 1916}

Noether's original manuscript no longer survives, but fortunately R. J. Humm made 
 a partial transcription, probably in early 1918. He also included the original  pagination, which indicates that his manuscript begins with page 15 of her text. Since a number of steps in Noether's arguments are based on equations from the first 14  pages, any attempt to reconstruct how she obtained these results would be necessarily  conjectural. Here I will simply take  such claims as established facts;  I will follow the same procedure when Noether draws on results in \cite{Hilbert1915} and \cite{Ein-6}. 
 By so doing, the general  train of her arguments is not difficult to follow. They show that Hilbert's energy vector as well as Einstein's pseudotensor representing gravitational energy can both be decomposed into two parts, one of which will have vanishing divergence, whereas the other vanishes as a result of the field equations.
Her analysis draws closely on Hilbert's own techniques in
\cite{Hilbert1915}, which she then applies in order to analyze Einstein's construct in \cite{Ein-6}.

Noether's analysis of Hilbert's energy vector exploited the fact that his ``world
function'' takes the form $H=K+L$ and that $K$ is defined solely by the metric tensor and its first and second derivatives. 
In her manuscript, 
she employs notation that deviates only slightly from that found in the two papers she discusses. 
For \cite{Hilbert1915} she begins her discussion of Hilbert's energy vector (\ref{eq:Hilbert-vector})  by looking at the vacuum case, $H=K$, where the last two terms $c^l=d^l=0$, since these only enter through the electromagnetic potential. She proceeds then to produce a decomposition of Hilbert's expression into a sum of two vectors, one of which vanishes by virtue of the field equations, whereas the divergence of the other vanishes identically, i.e., independent of the field equations.

Hilbert writes $p_s^i$ for $\frac {\partial p^i}{\partial w_s}$, and 
for the Lie variation:
\begin{equation}\label{eq:Lie-var}
\delta g^{\mu\nu} \equiv  p^{\mu\nu}= \sum_s (g^{\mu\nu}_s p^s - g^{\mu s}p^{\nu}_s -g^{\nu s}p^{\mu}_s).
\end{equation}
Noether follows Hilbert's Theorem III,  writing:
\begin{equation}\label{eq:i_s}
i_s= \sum_{\mu\nu} [ \sqrt g K]_{\mu\nu} g^{\mu\nu}_s
\end{equation}
\begin{equation}\label{eq:i_s^l}
i_s^l= -2 \sum_{\mu} [ \sqrt g K]_{\mu s} g^{\mu l} ,
\end{equation}
and then noting that
\begin{equation}\label{eq:Hilbert-ident-1}
\frac{1}{\sqrt g }
\sum_{\mu\nu} [ \sqrt g K]_{\mu\nu} p^{\mu\nu} =\frac{1}{\sqrt g }
\sum_{sl} {i_s \,p^s+ i_s^l\, p^s_l}.
\end{equation}

Hilbert's Theorem III asserts that $i_s = \sum_{l}\frac{\partial i_s^l}{\partial w_l },$\footnote{\cite[895]{Renn/Stachel2007} note that Theorem III ``corresponds to the contracted Bianchi identities,'' an insight that Hilbert and his contemporaries failed to notice, although the Bianchi identities had been discovered decades earlier. For the story of their recovery in the context of general relativity, see \cite[263--272]{Rowe2018}.} which means that $i_s$ can be written as a divergence or expressed in the form of the identity:
\begin{equation}\label{eq:Hilbert-ident-2}
\sum_{\mu\nu} [ \sqrt g K]_{\mu\nu} g^{\mu\nu}_s + 2\sum_{l}\frac{ \partial ([ \sqrt g K]_{\mu s} g^{\mu l})}{\partial w_l} =0.
\end{equation}
Noether exploits this in showing that the left side of (\ref{eq:Hilbert-ident-1}) can be written as a divergence. To do this she 
introduces  the vector
\begin{equation}\label{eq:Noether-vector}
i^l= \sum_{s}\frac{i_s^l}{\sqrt g }p^s,
\end{equation}
in order to rewrite equation (\ref{eq:Hilbert-ident-1}) as

\begin{equation}\label{eq:Noether-ident-1}
\frac{1}{\sqrt g }
\sum_{\mu\nu} [ \sqrt g K]_{\mu\nu} p^{\mu\nu} = \frac{1}{\sqrt g }
\sum_{sl}( \frac{\partial{i_s^l}}{\partial{w_l}} \,p^s+ i_s^l\, p^s_l)= Div(\sum_s\frac{i_s^l}{\sqrt g} \,p^s)=
Div (i^l).
\end{equation}
Drawing on previous calculations, she asserts that for 
$e^l= K\, p^l-a^l-b^l$
\begin{equation}\label{eq:Div(e^l)}
\frac{1}{\sqrt g }
\sum_{\mu\nu} [\sqrt g K]_{\mu\nu} p^{\mu\nu} = Div(e^l).
\end{equation}
It follows from equations (\ref{eq:Noether-ident-1}) and (\ref{eq:Div(e^l)})
that $Div (e^l)= Div(i^l)$ and furthermore that 
 $Div (e^l-i^l)=0$ holds identically. By virtue of the fundamental equations $[\sqrt g K]_{\mu\nu}=0= Div (i^l)$ for arbitrary $p^s$, whereas (\ref{eq:i_s^l}) shows 
 that 
$i_s^l$ also vanishes, which means that by definition (\ref{eq:Noether-vector}) $i^l=0$.

From this, Noether concludes that in the vacuum case one can always decompose Hilbert's energy vector as:
\begin{equation}\label{eq:Noether-ident-2}
 e^l= i^l+(e^l-i^l),
\end{equation}
where the first part vanishes as a consequence of the fundamental equations $[\sqrt g H]_{\mu\nu} =0$, whereas the divergence of the second part vanishes identically. She then summarizes the physical significance of this result as follows:
``The energy is probably {\it not} to be regarded as a first integral (as in classical mechanics) because it contains the second derivatives of the $g^{\mu\nu}$, and these cannot be eliminated from the $a^l$ by means of the fundamental equations.''\footnote{Hilbert introduced $a^l$ in a purely formal manner; see the discussion of equation (\ref{eq:polar-g}) above.}
From here, Noether makes use of the identity $Div (e^l-i^l)=0$ to count the number of equations that the components of $e^l$ need to satisfy, arriving at 120 such conditions. She then shows that the identical argument goes through for the general Lagrangian $H$, so that Hilbert's energy vector can always be decomposed as above. 

Noether next takes up a similar analysis of Einstein's version of the energy laws in general relativity, published in \cite{Ein-6}, arriving at very similar results. She  begins by noting how Einstein bases his theory on the demand that the equations of motion be given by (\ref{eq:geodesic}). Noether then 
rewrites Einstein's matter-free field equations (\ref{eq:modular2}) with only two small notational differences: her spacetime coordinates appear as $w_{\alpha}$ instead of $x_{\alpha}$, and she suppresses the coefficient $-2\chi$ in the second equation, which Einstein introduced for physical reasons. Likewise, she writes Einstein's law for conservation of momentum and energy (\ref{energy conservation}) in the form
 $$\sum_l\frac{\partial({t_s^l+T_s^l})}{\partial{w_l}}=0.$$

Drawing on Hilbert's notation, and noting that for $\sqrt{-g}=1, \,\, H=K$, she writes for the Lagrangian derivative in (\ref{eq:modular2}):
\begin{equation}\label{eq:EN-modular2}
-[\sqrt g H]_{\mu\nu}= \sum_{\alpha}\frac{\partial}{\partial{w_{\alpha}}}\left\{\frac{\partial H}{\partial{g^{\mu\nu}_{\alpha}}}\right\} 
-\frac{\partial H}{\partial{g^{\mu\nu}}}.
\end{equation}
Noether now connects (\ref{eq:EN-modular2}) with Einstein's pseudotensor for gravitational energy $t^{\alpha}_{\sigma}$
in  (\ref{eq:modular3}).
Multiplying (\ref{eq:EN-modular2}) by $g^{\mu\nu}_{\sigma}$ and summing over the indices $\mu, \nu$ yields:
$$-\sum_{\mu\nu} g^{\mu\nu}_{\sigma}[\sqrt g H]_{\mu\nu}= \frac{\partial}{\partial{w_{\alpha}}}
\sum_{\mu\nu}g^{\mu\nu}_{\sigma}\frac{\partial H}{\partial{g^{\mu\nu}_{\alpha}}} 
-\frac{\partial H}{\partial{w_{\sigma}}},$$
and thus
\begin{equation}\label{eq:EN-modular3}
-\sum_{\mu\nu} g^{\mu\nu}_{\sigma}[\sqrt g H]_{\mu\nu}=
 \sum_{\alpha}\frac {\partial t^{\alpha}_{\sigma}}{\partial w_{\alpha}}
\end{equation}
in view of (\ref{eq:modular3}).

Using Hilbert's Theorem III and  the identity (\ref{eq:Hilbert-ident-1}), Noether next obtains:

\begin{equation}\label{eq:Noether-ident-3}
\sum_{\mu\nu} [ \sqrt g H]_{\mu\nu} p^{\mu\nu} =
\sum_{sl} \frac{\partial{i_s^l}}{\partial{w_l}} \,p^s+ \sum_{sl}i_s^l\, p^s_l,
\end{equation}
and in place of (\ref{eq:i_s^l}) she writes: 
\begin{equation}\label{eq:Noether-ident-4}
-2 \sum_{\mu} [ \sqrt g H]_{\mu s} g^{\mu l} = t^l_s + r^l_s.
\end{equation}
Her claim is that  $i_s^l  = t^l_s + r^l_s$ and that $Div(i_s^l)=Div(t^l_s)$, so that $Div(r^l_s)\equiv 0$.
She proves this by multiplying (\ref{eq:EN-modular3}) by $p^s$ and (\ref{eq:Noether-ident-4}) by $p^s_l$, and then adding these two equations to get:
\begin{equation}\label{eq:Noether-ident-5}
\frac{1}{\sqrt g }\sum_{\mu\nu} [ \sqrt g H]_{\mu\nu} p^{\mu\nu} = \sum_{l}\frac{\partial{t^l_s}}{\partial{w_l}}p^s+ \sum_{sl} (t^l_s + r^l_s)p^s_l.
\end{equation}
Comparing coefficients in (\ref{eq:Noether-ident-3}) and (\ref{eq:Noether-ident-5}), Noether deduces the equations:
\begin{equation}\label{eq:Noether-ident-6}
\sum_{l} \frac{\partial{i_s^l}}{\partial{w_l}}=\sum_{l} \frac{\partial{t_s^l}}{\partial{w_l}};\, i^l_s= t^l_s + r^l_s,
\end{equation}
from which follows that $$\sum_{l} \frac{\partial{r_s^l}}{\partial{w_l}}= Div(r_s^l) \equiv 0,$$ under the assumption that $\sqrt{-g}=1$ holds.

Summarizing, she concludes that the Einsteinian gravitational pseudo-tensor $t_s^l$ also decomposes into two parts,
$t^l_s = i_s^l  - r^l_s$, where by (\ref{eq:i_s^l}) $i_s^l$ 
 vanishes as a consequence of the field equations, whereas the divergence of  $r^l_s$ vanishes identically, i.e., independent of the field equations. 
Noether actually shows that $i^l_s= t^l_s + r^l_s$, the second equation in (\ref{eq:Noether-ident-6}), is equivalent to Einstein's field equations written in the form (\ref{eq:modular4}). Finally, she briefly notes that the same considerations hold in the presence of matter, just as in the case of Hilbert's theory.

Humm's copy of Noether's manuscript contains no date, so we can only fix bounds for the period during which she must have written it. In his correspondence with Einstein from late May and early June of 1916, Hilbert alluded to Noether's 
efforts to reconcile their approaches to energy laws in general relativity. Much later, in January 1918, Hilbert and Klein both made reference to the results she had obtained more than one year earlier, so probably by December 1916 at the latest. Her text, on the other hand, contains no mention of \cite{Ein-5}, which surely circulated in G\"ottingen soon after its publication in early November 1916. Had she known of this text at the time, Noether would have most likely referred to the arguments Einstein set forth therein. These circumstances suggest that she probably completed her manuscript between June and October of 1916. After this date, Einstein published several times on energy conservation,\footnote{In addition to \cite{Ein-5}, see \cite{Ein-9} and \cite{Ein-8}.} which proved to be one of the most hotly debated issues in his theory of gravitation. For the G\"ottingen reception of general relativity, however, the most important of these notes was \cite{Ein-5}, to which we now turn.

\section{Einstein and Weyl respond to Hilbert}

Einstein recognized the importance of deriving his  field equations for general relativity from an
appropriate variational principle, but he strongly opposed Hilbert's effort to link the new
 theory of gravitation with Mie's electromagnetic theory of matter.  He originally thought about addressing this issue in 
\cite{Ein-6}, which quickly came to be regarded as a canonical text for the theory \cite{Gutfreund/Renn2015}. Among Einstein's posthumous  papers, one finds  an unpublished appendix written for 
\cite{Ein-6}, in which Einstein adopts Hilbert's variational methods, but with a general matter tensor rather than Hilbert's $L$. In a footnote, he  criticizes Hilbert for adopting Mie's matter function, which was based, of course, on the electrodynamic variables alone \cite[346]{Ein-11}. Quite possibly, Einstein withdrew this part of the text so as to avoid any potential polemics. He may have also considered this issue too important to merely appear in an appendix, and so he decided instead to publish a separate note on this topic. In  
 late October 1916 he submitted \cite{Ein-5} for publication in the {\it Sitzungsberichte der
Preu\ss ischen Akademie}. This provides a much fuller account of methods for 
deriving the fundamental equations of general relativity using variational principles. In the introduction, he wrote:

\begin{quote}
The general theory of relativity has recently been given in a
particularly clear form by H.A. Lorentz and D. Hilbert, who have
deduced its equations from one single principle of variation. The
same thing will be done in the present paper. But my purpose here is
to present the fundamental connections in as perspicuous a manner as
possible, and in as general terms as is permissible from the point of
view of the general theory of relativity. In particular we shall make as
few specializing assumptions as possible, in marked contrast to Hilbert's
treatment of
the subject (\cite[165]{Ein-5}).

\end{quote}

 Einstein thus employed Lagrangian equations of the type Hilbert derived earlier, but he began with only some 
general assumptions about
such functions ${\frak H}$ of the field variables $g^{\mu\nu}, \, q_{\rho}$ and their derivatives. He first noted that 
 the second derivatives $g^{\mu\nu}_{\sigma\tau}$
in ${\frak H}$ could be removed by partial integration, leading to a new Lagrangian ${\frak H}^*$ which satisfies
$$\int {\frak H} d\tau = \int {\frak H}^* d\tau + F,$$ where
$F$ is a surface term that can be neglected when the integral is suitably varied.
In this way, Einstein was able to substitute ${\frak H}^*$ for ${\frak H}$ in his variational principle
\begin{equation}\label{eq:var-princ}
\delta \biggl \{\int {\frak H}\,d\tau \biggr \} = \delta \biggl \{\int {\frak H}^*\,d\tau \biggr \}=0.
\end{equation}
This leads to the Lagrangian equations:
\begin{equation}\label{eq:gen.Lagr.eqns1}
\sum_{\alpha}\frac{\partial}{\partial x_{\alpha}}\biggl (\frac{\partial {\frak H}^*}{\partial
g^{\mu\nu}_{\alpha}}\biggr ) - \frac{\partial {\frak H}^*}{\partial g^{\mu\nu}} = 0.
\end{equation}
\begin{equation}\label{eq:gen.Lagr.eqns2}
\sum_{\alpha}\frac{\partial}{\partial x_{\alpha}}\biggl (\frac{\partial {\frak H}^*}{\partial
q_{\rho\alpha}}\biggr ) - \frac{\partial {\frak H}^*}{\partial q_{\rho}} = 0.
\end{equation}

Einstein next  
assumed $\frak H$ can be written $\frak H = \frak G + \frak M$ in order to assert the separate existence of the
gravitational field from matter. Furthermore, he assumed that $\frak M$ was a
function of the four electrodynamic variables $q_{\rho}$, their derivatives $q_{\rho\alpha}$ and $g^{\mu\nu}$. His $\frak G$ took the form
${\frak G}(g^{\mu\nu}, g^{\mu\nu}_{\sigma}, g^{\mu\nu}_{\sigma\tau})$, where the
coefficients of the
$g^{\mu\nu}_{\sigma\tau}$ were linear in the $g^{\mu\nu}$. 
By introducing a  function
$\frak G^*$  analogous to $\frak H^*$, Einstein was able to
deduce general gravitational field
equations of the form
\begin{equation}\label{eq:gen.fd.eqns1}
\sum_{\alpha}\frac{\partial}{\partial x_{\alpha}}\biggl (\frac{\partial {\frak G}^*}{\partial
g^{\mu\nu}_{\alpha}}\biggr ) - \frac{\partial {\frak G}^*}{\partial g^{\mu\nu}} = \frac{\partial {\frak M}}{\partial g^{\mu\nu}}.
\end{equation}
\begin{equation}\label{eq:gen.fd.eqns2}
\sum_{\alpha}\frac{\partial}{\partial x_{\alpha}}\biggl (\frac{\partial {\frak M}}{\partial
q_{\rho\alpha}}\biggr ) - \frac{\partial {\frak M}}{\partial q_{\rho}} = 0.
\end{equation}

Einstein next proceeded to specify further assumptions of his theory. This required that
$$ds^2 = \sum_{\mu,\nu} g_{\mu\nu}dx_{\mu}dx_{\nu},\quad H= \frac{{\frak H}}{\sqrt {-g}}, \quad
G= \frac{{\frak G}}{\sqrt {-g}}, \quad M= \frac{{\frak M}}{\sqrt {-g}}$$ all be invariants
under general coordinate transformations. This placed only limited restrictions on the
matter fields, but $G$, up to a constant factor, had to be the Riemann curvature scalar, which 
entails that ${\frak G}^*$ must also be uniquely determined.\footnote{Einstein never cited a mathematical source for this and other related claims, though he was well aware that his whole theory depended on this uniqueness property (see \cite[167, footnote 1]{Ein-5}). Quite possibly this was part of ``folklore'' knowledge among experts on the Ricci calculus, in which case Einstein might have picked this up from Marcel Grossmann. In 1920 Hermann Weyl published a proof in an appendix to the fourth edition of \cite{Weyl-4} (see \cite{Weyl-5}), noting that the first proof was given by Hermann Vermeil in 1917; see also \cite[43]{Pau-2}.} 
In a footnote, Einstein gave an explicit formula for $\frak G^*$:
\begin{equation}\label{eq:G^*}
{\frak G^*} = \sqrt{-g}\sum_{\alpha\beta\mu\nu}g^{\mu\nu}(\Gamma^{\alpha}_{\mu\beta}\Gamma^{\beta}_{\nu\alpha}-\Gamma^{\alpha}_{\mu\nu}\Gamma^{\beta}_{\alpha\beta}).
\end{equation}
This generalizes the Lagrangian (\ref{eq:modular-H}) that Einstein used in
 \cite{Ein-6}. In his letter to Weyl, cited above, Einstein repudiated (\ref{eq:modular-H}), noting that $\frak G^*$ is the required gravitational Lagrangian for generally covariant field equations, as Hilbert had shown.

Einstein then went on to carry out the variation $\int {\frak G}^*\,d\tau$, followed by the usual partial integrations, from which he
 deduced
four identities ($\sigma = 1,2,3,4$):
\begin{equation}\label{eq:4identities}
\sum_{\nu\alpha}\frac{\partial^2}{\partial x_{\nu}\partial x_{\alpha}}\biggl (\sum_{\mu}g^{\mu\nu}\frac{\partial {\frak G}^*}{\partial
g^{\mu\sigma}_{\alpha}}\biggr ) \equiv 0.
\end{equation}
From the general field equations (\ref{eq:gen.fd.eqns1}) he then derived
\begin{equation}\label{eq:AE-energy1}
\sum_{\alpha}\frac{\partial}{\partial x_{\alpha}}\biggl (\sum_{\mu}g^{\mu\nu}\frac{\partial {\frak G}^*}{\partial
g^{\mu\sigma}_{\alpha}}\biggr ) = - ({\frak T}_{\sigma}^{\nu} + {\frak t}_{\sigma}^{\nu}),
\end{equation}
where the terms on the right side of the equations denote
$${\frak T}_{\sigma}^{\nu} = - \sum_{\mu}\frac{\partial {\frak M}}{\partial g^{\mu\sigma}}g^{\mu\nu}; \,\,
 {\frak t}_{\sigma}^{\nu} = \frac{1}{2}\biggl ({\frak G}^*\delta_{\sigma}^{\nu}
-\sum_{\mu\alpha} \frac{\partial {\frak G}^*}{\partial g^{\mu\alpha}_{\nu}}g^{\mu\alpha}_{\sigma}\biggr ).$$

From (\ref{eq:AE-energy1}) and the identities (\ref{eq:4identities}), Einstein could now deduce his version of the conservation laws:
\begin{equation}\label{eq:AE-energy2}
 \sum_{\nu}\frac{\partial}{\partial x_{\nu}}({\frak T}_{\sigma}^{\nu} + {\frak t}_{\sigma}^{\nu}) =0.
\end{equation}
As before, he designated the ${\frak T}_{\sigma}^{\nu}$ as the
energy components of matter, whereas
the ${\frak t}_{\sigma}^{\nu}$ he regarded as
the components of the gravitational energy.
In closing, he derived the four equations for the energy components of matter
\begin{equation}\label{eq:AE-energy3} 
 \sum_{\mu\nu}\frac{\partial{{\frak T}_{\sigma}^{\nu}}}{\partial x_{\nu}}+ \frac{1}{2}g^{\mu\nu}_{\sigma}{\frak T}_{\mu\nu} =0.
\end{equation}
Einstein emphasized that in deriving the  conservation laws (\ref{eq:AE-energy2}) and
(\ref{eq:AE-energy3}) he needed only the gravitational field equations (\ref{eq:gen.fd.eqns1}) but not the
field equations for matter  
(\ref{eq:gen.fd.eqns2}).

Readers familiar with \cite{Hilbert1915} surely recognized 
Einstein's desire to place
his variational approach to the fundamental equations of his gravitational theory
 in the sharpest possible contrast with Hilbert's. He had struggled during the late spring of 1916 to understand how Hilbert constructed his invariant energy vector, but openly admitted that its physical significance eluded him entirely.
Apparently he felt no differently one year later when he spoke about it with Rudolf Humm. He wondered how energy could be a vector, but also what sense it made when its very definition was multi-valued, since it contained an arbitrary vector
\cite[70]{Rowe2019}. 

Einstein also had deep misgivings about Hilbert's methodological approach.
Writing to Hermann Weyl shortly after \cite{Ein-5} was published, he confessed:
\begin{quote}

To me Hilbert's {\it Ansatz} about matter appears to be childish, just like an infant who is unaware
of the pitfalls of the real world\dots . In any case, one cannot accept the mixture of well-founded
considerations arising from the postulate of general relativity and unfounded, risky hypotheses about the structure of the electron\dots . I am the first to admit that the discovery of the proper hypothesis, or the Hamilton function, of the structure of the electron is one of the most important tasks of the current theory. The ``axiomatic method'', however, can be of little use in
this. (Einstein to Weyl, 23 November 1916, \cite[366]{Ein-11}.)
\end{quote}

Einstein's letter was written in response to a draft of \cite{Weyl-1}, which employed variational methods to deduce conservation laws
in general relativity.
Weyl shared Einstein's criticism of  Hilbert's theory, especially its reliance on Mie's theory and the assumption of the special matter tensor
$T_{\mu\nu} = \frac{\partial  L}{\partial g^{\mu\nu}}.$ He thus 
emphasized the
provisional nature of all efforts to base gravitational theory on variational principles owing to lack of knowledge about elementary particles. ``Under these
circumstances,'' he wrote, ``it appears to me important to
formulate {\it a Hamiltonian principle that carries as far as our present
knowledge of
matter reaches} \dots '' (\cite[118]{Weyl-1}).

Weyl's theory combined a general matter function to the gravitational and electromagnetic fields. The field effects are then measured by the action integrals: $$\int H\, d\omega, \,\, \int L\, d\omega,$$
where $H$ is given by (\ref{eq:G^*}) and $$L= \frac{1}{2}F_{ik}F^{ik} = \frac{1}{2}g^{ij}g^{kh}F_{ik}F_{jh},
\quad F_{ik}= \frac{\partial \phi_k}{\partial x_i}- \frac{\partial \phi_i}
{\partial x_k}.$$ Alongside these field actions, Weyl introduces analogous substance actions given by integrals based on density functions for matter $dm$ and electricity $de$: $$\int \biggl \{dm \int \sqrt{g_{ik} \,dx_idx_k} \biggr \}, \,\, 
\int \biggl \{de \int \phi_i \, dx_i \biggr \}.$$
All of these ingredients enter into Weyl's ``world function'' $F$ defined on a given region $\Omega$, for
which he postulates that under variations of the field variables that vanish at the boundary of $\Omega$ and infinitesimal spacetime displacements of the substance elements this $F$ will be an extremum. 

 From this postulate, he immediately derives corresponding results for gravitation, electromagnetism, and mechanics. Thus, by varying the $g^{ij}$ while holding the $\phi_i$ and the worldlines of substance fixed, one gets Einstein's gravitational equations (\ref{eq:einst}). Varying the $\phi_i$ yields the Maxwell-Lorentz equations
$$\frac{1}{\sqrt g}\frac{\partial(\sqrt g F^{ik})}{\partial x_k} = J^i=\epsilon \frac{dx_i}{ds}.$$ Finally, varying the worldlines of the substance elements leads to the equations of motion for mass points when acted on by electromagnetic forces
 \begin{equation}\label{eq:geodesic-forces}
\rho \biggl (\frac{d^2x_i}{ds^2} - \Gamma^i_{hk}\frac{dx_h}{ds}\frac{dx_k}{ds} \biggr ) =p^i.
\end{equation}
Here the $p^i$ are the contravariant components of the force corresponding to the covariant $$p_i= \sum_k F_{ik}J^k.$$ Weyl remarks further that (\ref{eq:geodesic-forces}) can be shown to follow directly from the other two systems of field equations. He regarded these findings as purely phenomenological deductions analogous to those of classical Hamiltonian mechanics. 
 This approach thus stressed flexibility, and 
the following year he 
elaborated on some of these ideas in the first edition of {\it Raum--Zeit--Materie} \cite{Weyl-4}.

In section 2 of \cite{Weyl-1}, he introduces a general action integral defined on a region of spacetime $\Omega$ for which $$\int_{\Omega} (H-M)\, d\omega$$ is an extremum. Here the matter-density action $M$ is closely related to Einstein's energy-momentum tensor $T_{ik}$; the latter is defined, however, in connection with the total derivative of the former. Weyl's objective is to deduce Einstein's
energy-momentum equations for matter (\ref{eq:AE-energy3}) by an appropriate variation applied to ${\frak M}= M \sqrt g$.
He was apparently the first author to emphasize that the conservation of energy-momentum in general relativity should be deduced from a variational principle under which the variation of the field quantities is induced by coordinate transformations. In Weyl's language, the field variables are ``mitgenommen'' by means of infinitesimal coordinate transformations \cite[117]{Weyl-1}. One year later, Klein and also Noether alluded to earlier work of Sophus Lie, who introduced this method in his new group-theoretic approach to differential equations (see \cite{Haw}). Within the context of the calculus of variations, this technique came to be known as Lie variation. As was pointed out by Janssen and Renn, Einstein only gradually came to appreciate the importance of this mathematical technique for field physics (\cite[863]{Janssen/Renn2007}). 

Adopting Weyl's notation, one considers transformations
$$ x_i \rightarrow x_i+\epsilon\xi_i(x_1,x_2,x_3,x_4)$$ for
infinitesimal $\epsilon$ and $\xi_i$, which along with
their derivatives vanish on the boundary of integration, and then calculates
 $\delta g^{ik}$, the induced variation of the field quantities: 
$$\delta g^{ik} = \epsilon( g^{\alpha k}\frac{\partial\xi_i}{\partial x_{\alpha}} + g^{i\beta} \frac{\partial\xi_k}{\partial x_{\beta}}).$$ 
Weyl then distinguished this $\delta$-variation from a second $\Delta$-variation given by 
 \begin{equation}\label{eq:Delta-var}
\Delta g^{ik} = \delta g^{ik} -\epsilon
\frac{\partial g^{ik}}{\partial x_{\alpha}}\xi_{\alpha}.
 \end{equation} 
Under a $\Delta$-variation, the domain of 
definition $\Xi$ for the coordinates $(x_1,x_2,x_3,x_4)$ corresponding to the 
region $\Omega$ remains identical, leading to what Weyl calls a 
 virtual displacement. 

Using this
variational technique, Weyl rederives Einstein's
energy-momentum equations for matter (\ref{eq:AE-energy3}) (\cite[124]{Weyl-1}).
Writing $dx$ for $dx_1dx_2dx_3dx_4$, he notes that $\int {\frak M}\, dx$ is an invariant and that
$$\int_{\Xi} \Delta {\frak M}\, dx =0.$$ Furthermore,
$$\int_{\Xi} \Delta {\frak M}\, dx = \int {\frak T}_{ik}\,\Delta g^{ik}\, dx, \quad{\frak T}_{ik} = \sqrt g T_{ik}.$$
Substituting (\ref{eq:Delta-var}) and carrying out the partial integration leads to
$$ \int \biggl \{ \sum_{krs}\frac{\partial{{\frak T}_i^k}}{\partial {x_k}}+ \frac{1}{2}\frac{{\partial {g^{rs}}}}{{\partial {x_i}}}{\frak T}_{rs}\biggr \}\,\xi_i dx =0,$$
and since $\xi_i$ is arbitrary, he gets (\ref{eq:AE-energy3}):
$$\sum_{krs}\frac{\partial{{\frak T}_i^k}}{\partial {x_k}}+ \frac{1}{2}\frac{{\partial {g^{rs}}}}{{\partial {x_i}}}{\frak T}_{rs}=0.$$

Weyl next points out that a parallel argument using the gravitational action $H$ leads to four analogous equations satisfied by the Einstein tensor
$R^{\mu\nu}- \frac{1}{2}g^{\mu\nu}R$. Written in modern notation, these are $$(R^{\mu\nu}- \frac{1}{2}g^{\mu\nu}R)_{;\nu}=0,$$
known today as contracted
Bianchi identities. Since these are formally identical to the equations (\ref{eq:AE-energy3}), 
Weyl made the noteworthy observation that the latter equations are an immediate consequence of Einstein's field equations (\ref{eq:einst}), written in the form
\begin{equation}
R^{\mu\nu}- \frac{1}{2}g^{\mu\nu}R = -\kappa T^{\mu\nu}.
\end{equation}
He further observed that this was a natural consequence of a generally covariant theory, since the freedom to choose any coordinate system is reflected in the fact that these ten gravitational field equations satisfy four differential identities. 
Although Weyl clearly recognized the connection between his results and Hilbert's Theorem I, he made no direct comments about the latter. Instead, he cited Hilbert's second note \cite{Hilbert1917}, which addressed the problem of causality in GR while proposing a method for handling Cauchy problems. 
 Nor did he draw any clear distinction between relativistic conservation laws and their counterparts in classical mechanics.
Working within this novel context, Weyl's focus was on adapting variational principles to the new field physics, following the lead of Hilbert, Lorentz, and Einstein. None of these mathematicians and physicists was deeply versed in the fine points of Ricci's tensor calculus, including the full Bianchi identities.\footnote{As was pointed out in \cite[274--276]{Pais}; for the ensuing history, see \cite[263--272]{Rowe2018}.} Due to this circumstance, they came to regard the contracted Bianchi identities as a result obtained by using variational methods.

\section{Klein's Critique of \cite{Hilbert1915} }

The papers by Einstein and Weyl discussed above were carefully studied by
Felix Klein, who from early 1917 began 
 to play an active
role in ongoing discussions of conceptual problems in general relativity. 
As noted above in section 3, in January 1918  Klein and Hilbert reached a first consensus
regarding some fundamental issues related to general relativistic physics \cite{Kl-1}.
With regard to variational methods and conservation laws derived from them, Klein
emphasized the importance of separating  formal deductions from 
 physical claims, such as those that form the basis for Einstein's new gravitational theory. Much of what he and Hilbert discussed centered on the distinction between theories based on invariants of the orthogonal group and those that arise from a variational problem based on general invariants, as in Hilbert's adaptation of Einstein's theory.

Klein introduced a special Lagrangian in place of $L$, namely
$$L=\alpha Q=-\alpha\sum_{\mu\nu\rho\sigma}(q_{\mu\nu}-q_{\nu\mu})(q_{\rho\sigma}-q_{\sigma\rho})(g^{\mu\rho}g^{\nu\sigma}-g^{\mu\sigma}g^{\nu\rho}),$$
where $-\alpha$ is Einstein's $\kappa = \frac{8\pi K}{c^2}$ and $K$ the universal gravitational constant from Newton's theory \cite[333]{Ein-6}. He then observes that the tiny value $-\alpha = 1.87\cdot 10^{-27}$ will ensure that the new theory accords with Maxwell's theory, for which $\alpha =0$. Klein next takes the two integrals separately:
$$I_1= \int K d\omega, \,\, I_2= \alpha \int Q d\omega,$$ and carries out the variation in a purely formal manner, writing: $$\delta I_1 = \int K_{\mu\nu}\delta g^{\mu\nu}d\omega;$$
$$\delta I_2 = \alpha \int (\sum_{\mu\nu}Q_{\mu\nu}\delta g^{\mu\nu}+\sum_{\rho}Q^{\rho}
 \delta q_{\rho})d\omega.$$ Here $K_{\mu\nu}$ is Hilbert's $[\sqrt g K]_{\mu\nu} : \sqrt g$, whereas
$Q_{\mu\nu}=(\frac{\partial \sqrt g Q}{\partial g^{\mu\nu}})    : \sqrt g$, and the vector
$$Q^{\rho}=-\sum_{\sigma}\frac{\partial(\frac{\partial \sqrt g Q}{\partial q^{\rho\sigma}})}{\partial w^{\sigma}} : \sqrt g.$$ Clearly the $Q_{\mu\nu}$ are the coefficient's in (\ref{eq:Hilbert-EM-Energie}), Hilbert's expression for electromagnetic energy, so Klein called these the energy components of the electromagnetic field. He further identified 
$Q^{\rho}=0$ as the counterpart to the Maxwell equations.

Carrying out the variation for $I_1$ leads almost immediately to the four differential equations that Hilbert had derived using Theorem III (see (\ref{eq:Hilbert-ident-2})):
 \begin{equation}\label{Klein-identities-1}
\sqrt g \sum_{\mu\nu}K_{\mu\nu} g^{\mu\nu}_{\sigma}+ 2\sum_{\mu\nu}\frac{\partial(\sqrt g K_{\mu\sigma} 
g^{\mu\nu})}{\partial w^{\nu}}
=0, \,\, \sigma = 1,\,2,\,3,\,4,
 \end{equation}
which Klein summarizes in the statement that the vectorial divergence of $K_{\mu\nu}$ vanishes. For the variation of 
$I_2$ he obtains:
 \begin{equation}\label{Klein-identities-2}
\sum_{\mu\nu}(\sqrt g Q_{\mu\nu} g^{\mu\nu}_{\sigma}+ 2\sum_{\mu\nu}\frac{\partial\sqrt g (Q_{\mu\sigma} 
g^{\mu\nu})}{\partial w^{\nu}})+\sum_{\rho}(\sqrt g Q^{\rho} (
q_{\rho\sigma}-q_{\sigma\rho}))
=0, \,\, \sigma = 1,\,2,\,3,\,4.
 \end{equation}
Only at this point does Klein make use of the field equations, which here appear in the form:
 $$K_{\mu\nu}+\alpha Q_{\mu\nu}=0; \,\,Q^{\rho}=0.$$ Multiplying (\ref{Klein-identities-2}) by $\alpha$ and adding this to (\ref{Klein-identities-1}) yields:
	 \begin{equation}\label{Klein-identities-3}
\sum_{\mu\nu}\sqrt g ( K_{\mu\nu}+\alpha Q_{\mu\nu}) g^{\mu\nu}_{\sigma}+ 
2\sum_{\mu\nu}\frac{\partial(\sqrt g ( K_{\mu\sigma} +\alpha Q_{\mu\nu})
g^{\mu\nu})}{\partial w^{\nu}}
+\alpha \sum_{\rho}(\sqrt g 
Q^{\rho} (q_{\rho\sigma}-q_{\sigma\rho}))
=0.
 \end{equation}
From the equations (\ref{Klein-identities-3}) Klein immediately deduces that the four equations $Q^{\rho}=0$ follow directly from the ten equations 
$K_{\mu\nu}+\alpha Q_{\mu\nu}=0$. If, on the other hand, one takes the generalized Maxwell equations $Q^{\rho}=0$  
alongside the four identities (\ref{Klein-identities-2}), then one can conclude that the energy components $Q_{\mu\nu}$ have a vanishing vectorial divergence.

This straightforward analysis pointed to one of the glaring weaknesses in 
 Hilbert's theory,  namely the
use he made  of Theorem 1 to deduce 
 four identities from his fourteen fundamental equations. Hilbert's idea of reducing electrodynamics to gravitational effects hinged on applying Theorem 1 to the world function $H=K+L$. What 
Klein simply pointed out was that by handling gravity and
electromagnetism separately, one can derive four identities
from each, namely the four Lagrangian equations
derivable from $\delta\int K d\omega =0$ and $\delta\int Q d\omega =0$,
respectively. This meant that  
Hilbert's Theorem I led to {\it eight}
identities and not just four, an observation that  seriously undermined
 his unification program.

Klein was also able to shed new light on Hilbert's invariant energy vector $e^{\nu}$ by slightly transforming equations (\ref{Klein-identities-3}). This led to the recognition that $e^{\nu}$ could be decomposed into a sum of two vectors, the first being
$$e_1^{\nu}= -2\sum_{\mu\sigma} (( K_{\mu\sigma}+\alpha Q_{\mu\sigma}) g^{\mu\nu}+\frac{\alpha}{2}Q^{\nu}q_{\sigma})p^{\sigma}$$
and the second $e_2^{\nu}$  having vanishing divergence.\footnote{Klein also noted certain properties of $e_2^{\nu}$, but he found it too difficult to calculate directly. A few months later he discovered a different way to derive Hilbert's $e^l$ and presented this in \cite{Kl-2}.} Since the first vector vanishes by virtue of the field equations, Klein concludes that Hilbert's invariant energy theorem (\ref{eq:Hilbert-energy}) is merely an identity and thus by no means analogous to conservation of energy in classical mechanics. These findings were clearly in accord with what Emmy Noether had already pointed out to Hilbert more than a year before. Since she still had her manuscript, she was able to show her derivation to Klein. This probably  took place on or shortly after 22 January 1918, when he spoke about these matters at a meeting of the G\"ottingen Mathematical Society. After mentioning her earlier results, Klein somewhat dismissively wrote that she had not brought out their importance as decisively as he had done in his lecture \cite[559]{Kl-1}.

\section{Klein's Correspondence with Einstein on Energy Conservation}

 Klein's open letter to Hilbert contained similar remarks about Einstein's derivation of the
``conservation laws'' (\ref{eq:AE-energy2}) in \cite{Ein-5} (the quotation marks are Klein's).
 Klein claimed that these
 four equations should also be regarded as mathematical identities, by which he apparently meant that they were consequences of the field equations. This assertion was disputed by Einstein and led to
some lengthy exchanges between him and Klein during the month of March 1918. 
On 13 March, Einstein wrote him: 
\begin{quote}
It was with great pleasure that I read your extremely clear and elegant explanations regarding Hilbert’s first note. However, I consider your remark about my formulation of the conservation laws to be inaccurate. For equation [(\ref{eq:AE-energy2})] is by no means an identity, any more than [(\ref{eq:AE-energy1})]; only [(\ref{eq:4identities})] is an identity. The conditions [(\ref{eq:AE-energy1})] are the mixed form of the field equations of gravitation. [(\ref{eq:AE-energy2})] follows from [(\ref{eq:AE-energy1})] on the basis of the identity [(\ref{eq:4identities})]. The relations here are exactly analogous to those of nonrelativistic theories. \cite[673]{Einstein1998b}
\end{quote}

Einstein might have noticed that what Klein meant by an identity differed from his own understanding, but he was mainly intent on spelling out the physical importance of the pseudo-tensor
 ${\frak t}_{\sigma}^{\nu}$
 in the conservation laws (\ref{eq:AE-energy2}) The ${\frak t}_{\sigma}^{\nu}$'s not only lead to these laws but also with (\ref{eq:AE-energy1}) they provide a physical 
 interpretation entirely analogous to Gauss's law in electrostatics.

\begin{quote}

In the static case the number of ``lines of force'' running from a physical system to infinity is, according to [(\ref{eq:AE-energy1})], only dependent on the 3-dimensional spatial integrals 
$$\int ({\frak T}_{\sigma}^{\nu} + {\frak t}_{\sigma}^{\nu})dV$$
to be taken over the system and the gravitational field belonging to the system. This state of affairs can be expressed in the following way. As far as its gravitational influence at a great distance is concerned, any (quasi-static) system can be replaced by a point mass. The gravitational mass of 
this point mass is given by
$$\int ({\frak T}_{4}^{4} + {\frak t}_{4}^{4})dV$$
i.e., by the total energy (more precisely, total ``rest energy'') of the system, exactly as the inertial mass of the system. \dots

From [(\ref{eq:AE-energy2})] it can be concluded that the same integral 
$\int ({\frak T}_{4}^{4} + {\frak t}_{4}^{4})dV$
also determines the system's inertial mass. Without the introduction and interpretation of ${\frak t}_{\sigma}^{\nu}$, one cannot see that the inertial and gravitational mass of a system agree.

I hope that this anything but complete explanation will enable you to guess what I mean. Above all, though, I hope you will abandon your view that I had formulated an identity, that is, an equation that places no conditions on the quantities in it, as the energy law. \cite[674]{Einstein1998b}
\end{quote}

Regarding this last point, Klein was still thoroughly unpersuaded, and so he sent Einstein 
his ``rebuttal'' in a long letter from 20 March \cite[685--688]{Einstein1998b}. Klein's key assertion was that the equations (\ref{eq:AE-energy2}) are completely equivalent to $$\sum_{\nu}\frac{\partial ( K_{\sigma}^{\nu} +\alpha Q_{\sigma}^{\nu})}
{\partial w^{\nu}}=0$$ and that the latter are ``physically contentless.'' He meant by this nothing more than the observation that Einstein's conservation laws followed directly from the gravitational field equations. 

Klein further informed Einstein that 
Carl Runge had found a way to particularize 
the coordinate system to obtain conserved quantities directly from:
$$\sum_{\nu} \frac{\partial T^{\nu}_{\sigma}}{\partial x_{\nu}} =0.$$
 Delighted by this apparent breakthrough (``the pure egg of Columbus''), he
was anxious to learn what Einstein thought about Runge's finding. 
Emmy Noether already knew about this proposal, and she was highly skeptical. She was visiting her father in Erlangen, so Klein mailed her a draft of \cite{Kl-1}.along with a description of Runge's result. She quickly went to work and found from concrete examples that 
Runge's coordinate transformation led
to well-known identities that cannot be interpreted as energy
laws.\footnote{E. Noether to F. Klein, 12 March 1918, Nachlass Klein, (SUB), G\"ottingen.}

Einstein clarified his views on these matters in a letter from 24 March. In this reply, he 
underscored that the equations above 
contained {\it part} of the content of the field equations $$ K_{\sigma}^{\nu} +\alpha Q_{\sigma}^{\nu}=0.$$
The same was true for the equations 
$$\sum_{\nu}\frac{\partial ({\frak T}_{\sigma}^{\nu} + {\frak t}_{\sigma}^{\nu})}
{\partial x_{\nu}}=0,$$ though with the important advantage that these equations can be used to obtain an integral formulation for energy conservation on regions of space-time over which the ${\frak T}$'s and ${\frak t}$'s vanish. One then obtains $$\frac{d}{dx_4}\{\int ({\frak T}_{\sigma}^{4} + {\frak t}_{\sigma}^{4}) \}=0.$$
Einstein emphasized that ``the temporal constancy of these four integrals is a nontrivial consequence of the field equations and can be looked upon as entirely similar and equivalent to the momentum and energy conservation laws in the classical mechanics of continua'' \cite[697]{Einstein1998b}.

As for Runge's proposal for obtaining the conservation laws by 
particularizing the coordinate system, Einstein reported that he had explored that idea himself, 
 but had
given it up ``because the theory predicts energy losses
due to gravitational waves'' and these loses could not be taken
into account. Einstein included an offprint of his recent paper \cite{Ein-9}, in which he introduced the quadrupole formula for the propagation of gravitational radiation. This was a typical instance showing how Einstein could quickly cast aside a
mathematical idea when he noticed that it failed to conform to his physical understanding. Emmy Noether's reservations regarding Runge's approach were, of course, based on essentially mathematical considerations. Klein and Runge soon hereafter dropped this line of investigation, but Klein continued to explore the mathematical underpinnings of energy conservation in the context of invariant variational principles.

In mid-July, he wrote to Einstein with news of a first breakthrough: ``I have succeeded in finding the organic law of construction for Hilbert's energy vector'' \cite[833]{Einstein1998b}. Klein's innovation 
was surprisingly simple. Previously, he and others has carried out  infinitesimal variations using a vector field 
 $p^{\tau}$, which along with its derivatives was required to vanish on the boundary of the integration domain. Klein now dropped this restriction, so that in carrying out the variation he obtained an additional triple integral of the form
$$\int\int\int\sqrt g\{e^1dw^2dw^3dw^4+\dots +e^4dw^1dw^2dw^3\}.$$ He then found that Hilbert's energy vector was essentially identical to $(e^1,e^2,e^3,e^4)$,
differing only by terms with vanishing divergence.
 In his letter to Einstein, Klein reported that he hoped now to find his way to Einstein's formulation of energy conservation based on ${\frak T}^{\nu}_{\sigma} +{\frak t}^{\nu}_{\sigma}.$
Einstein answered: ``It is very good that you want to clarify the formal significance of the ${\frak t}^{\nu}_{\sigma}$. For I must admit that the derivation of the energy theorem for field and matter together appears unsatisfying from the mathematical standpoint, so that one cannot
characterize the ${\frak t}^{\nu}_{\sigma}$ formally'' \cite[834]{Einstein1998b}. Einstein was also unhappy about the fact that his pseudotensor was unsymmetric, unlike the matter tensor.\footnote{In 1951 Landau and Lifschitz introduced a symmetric pseudotensor for gravitational energy; unlike the Einstein pseudotensor, it conserves angular momentum.} 

It should be emphasized that Klein was working closely with 
Emmy Noether during this period, as he acknowledged in \cite{Kl-1} and \cite{Kl-2}. In fact, the latter paper and \cite{Noether1918b} should be seen as complementary studies, and in today's world would surely have been co-authored publications. On Monday, 22 July, Klein spoke on ``Hilberts Energievektor'' before the G\"ottingen Mathematical Society, one day before Noether's talk on ``Invariante Variationsprobleme.'' Klein then submitted the preliminary version of her findings to
 the G\"ottingen Scientific Society on Friday, 26 July, having done the same one week earlier with his manuscript for \cite{Kl-2}. Both papers underwent final revision in September and appeared in the {\it G\"ottinger Nachrichten} shortly afterward. 

By this time, H.A. Lorentz had also derived differential equations for energy conservation in gravitational fields, so his was a third formulation in addition to those of Einstein and Hilbert. It seemed evident that these different versions must be somehow related, and Klein hoped to explain how. Noether's earlier work on the same question clearly helped to move this project forward. 
Klein's framework in \cite {Kl-2} extends the one he utilized in \cite {Kl-1}. He now begins with a general variational problem for a scalar function $K$ viewed as a function of $g^{\mu\nu}, g^{\mu\nu}_{\rho}, g^{\mu\nu}_{\rho\sigma}$ alone. In this general setting he derives a series of identities leading to what he calls the principal theorem, which he writes in the form 
\begin{equation}\label{eq:princ-theorem}
\sum_{\mu\nu}\sqrt g (K_{\mu\nu}g^{\mu\nu}_{\tau})\equiv 2\sum_{\sigma}
\frac{\partial \sqrt g U^{\sigma}_{\tau}}{\partial w^{\sigma}},
\end{equation}
 where $K_{\mu\nu}$ is the Lagrangian derivative. This identity effectively turns the four expressions on
 the left-hand side into what Klein calls elementary divergences because they only involve the first derivative of the $g^{\mu\nu}$. The right-hand side derives from the triple integral above, which Klein introduced in order to derive Hilbert's energy vector. In the previous derivations this expression simply vanishes due to the conditions imposed on the boundary of integration.

Klein gave a simple extension of equation (\ref{eq:princ-theorem}) after writing it in the abbreviated form:
$$\frak K_{\mu\nu}g^{\mu\nu}_{\tau}\equiv 2 
\frac{\partial \frak U^{\sigma}_{\tau}}{\partial w^{\sigma}}.$$ He then noted that the Lagrangian derivative of any elementary divergence $\frak D\frak i\frak v$
vanishes. So for any $\frak K^*= \frak K + \frak D\frak i\frak v$, the left-hand side will remain the same, and the theorem then reads: $$\frak K_{\mu\nu}g^{\mu\nu}_{\tau}\equiv 2 
\frac{\partial \frak U^{*\sigma}_{\tau}}{\partial w^{\sigma}}.$$

Only at this point does Klein take up analysis of these expressions as invariants of groups. Those deriving from the left-hand side then correspond to invariants under general coordinate transformations (or, as one would say today, arbitrary diffeomorphisms). The $U^{\sigma}_{\tau}$, resp. 
$U^{*\sigma}_{\tau}$, on the other hand, are only invariant under affine transformations. This was also the case with Einstein's pseudo-tensor, but Klein now underscored the key property that Einstein had already noted before, namely that these affine invariants enter into an equation that is valid in all coordinate systems. In the present case, this reads:
\begin{equation}\label{eq:A-beta}
\frac{\partial (\frak K^{\sigma}_{\tau}+\frak U^{\sigma}_{\tau})}{\partial w^{\sigma}}\equiv 0,
\end{equation}
 or from the extended theorem,
\begin{equation}\label{eq:A-gamma}
\frac{\partial (\frak K^{\sigma}_{\tau}+\frak U^{*\sigma}_{\tau})}{\partial w^{\sigma}}\equiv 0.
\end{equation}
These are evidently purely mathematical deductions valid for any invariant scalar function $K$. 

Klein next turns to physics, by introducing the field equations in the simplest form suitable for his purposes, writing: $$\frak K^{\sigma}_{\tau}-\chi \frak T^{\sigma}_{\tau}=0.$$ Substituting in 
(\ref{eq:A-beta}) and (\ref{eq:A-gamma}), leads to two forms of the conservation laws,
$$\frac{\partial (\frak T^{\sigma}_{\tau}+\frac{1}{\chi}\frak U^{\sigma}_{\tau})}{\partial w^{\sigma}}= 0,$$
$$\frac{\partial (\frak T^{\sigma}_{\tau}+\frac{1}{\chi}\frak U^{*\sigma}_{\tau})}{\partial w^{\sigma}}= 0.$$ The first of these reflects the form Lorentz derives, whereas Klein shows that Einstein's formulation (\ref{eq:AE-energy2}) conforms with 
the second. Analyzing Hilbert's energy vector led to additional complications, but the net result was the same: except for additional terms of no physical significance, its form was also of the second type.

Soon after Klein's paper \cite {Kl-2} on the differential form of the conservation
laws  came out in October, he sent a copy to Einstein.
The latter responded with enthusiasm:
``I have already studied your paper most thoroughly and with
true amazement. You have clarified this difficult matter fully.
Everything is wonderfully transparent'' \cite[917]{Einstein1998b}. He was particularly delighted that Klein had not rejected his controversial pseudo-tensor for gravitational energy. Only one question still bothered him: how can one prove that Hilbert's expression is truly a generally covariant vector?

Klein answered with a calculation, but Einstein found the argument behind it insufficient \cite[932]{Einstein1998b}. One week later, after consulting with his {\it Assistent} Hermann Vermeil, Klein sent Einstein a new calculation. He realized that the argument was anything but elegant, but was eager to learn what Einstein thought of it \cite[936--937]{Einstein1998b}. He received this immediate response:
\begin{quote}
Thank you very much for the transparent proof, which I understood completely. The fact that it cannot be realized without calculation does not detract from your overall investigation, of course, since you make no use of the vector character of $e^{\sigma}$.-- In the whole theory, one thing still disturbs me formally, namely, that $T{\mu\nu}$ must necessarily be symmetric but not $t{\mu\nu}$, even though both must enter equivalently in the conservation law. Maybe this disparity will disappear when ``matter'' is included, and not just superficially as it has been up to now, but in a real way in the theory. \cite[938]{Einstein1998b}.\footnote{Landau and Lifschitz introduced a symmetric pseudotensor for gravitational energy in 1951.}
\end{quote}

A few days later, Klein had the opportunity to discuss this problem with Emmy Noether, who explained that Hilbert had already alluded to a general method for proving that $e^{\sigma}$ transformed as a vector in \cite{Hilbert1915}. Klein immediately wrote to Einstein  with a sketch of the proof, which did not depend on special properties of $K$ \cite[942--943]{Einstein1998b}. Once again, Noether emerged as the real expert when it came to unpacking the mysteries surrounding Hilbert's energy vector. 

\section{Noether's Two Theorems}

This was also the case with Hilbert's Theorem I and its role in the formulation of
conservation laws.
Klein's main concern in \cite{Kl-1} was the status of conservation laws in general relativity. 
Contrary to Einstein, he distinguished sharply between these new findings and traditional   
 conservation laws in classical mechanics. The latter, he argued, cannot simply be deduced from a variational principle; for example, 
 one cannot derive $$\frac{d(T+U)}{dt}=0$$ without invoking
 specific physical properties or principles, such as Newton's law of motion. 
Klein attached great significance to this issue in part because he wanted to promote
ideas from his ``Erlangen Program'' \cite{Kl-4},  which he was adapting into a general
doctrine applicable to the new physics. Relativity theory, according to
Klein, should not be thought of exclusively in terms of two groups -- the
Lorentz group of special relativity and the group of continuous point
transformations of general relativity -- but rather should be
broadly understood as the invariant theory {\it relative to some given
group} that happens to be relevant to a particular physical theory.
This was the mathematical context
Klein had in mind when he emphasized the distinction between conservation laws
in classical mechanics, special relativity, and the general theory of relativity.\footnote{ Klein's articles on
relativity theory originally appeared in the {\it G\"ottinger Nachrichten}
as well, but in 1921 he republished them along with additional commentary
in the first volume of his
collected works. In doing so, he placed them in a special section
entitled ``Zum Erlanger Programm'' (\cite[I: 411--612]{Kl-GMA}).}

Hilbert not only agreed with Klein's assertion, he went even further by expressing the opinion that 
 the lack of analogy between classical
energy conservation and his own energy equation was a characteristic feature of general relativity.
In his own inimitable manner, he
 even claimed {\it one could prove a theorem}
effectively ruling out conservation laws for general transformations
analogous to those that hold for the transformations of the orthogonal group.
Klein replied by saying:
``It would interest me very much to see the mathematical
proof carried out that you alluded to in your answer'' \cite[565]{Kl-1}.
 Hilbert's conjecture was resolved  some months later when 
Emmy Noether published ``Invariante Variationsprobleme'' \cite{Noether1918b}.

Noether was in Erlangen around the time  Klein was
putting the last touches on \cite{Kl-1}. From there 
 she wrote him on 29 February
1918: ``I thank you very much for sending me your note and today's
letter [same day delivery was not uncommon in those times], and I'm very
excited about your second note \cite{Kl-2}; the notes will certainly contribute much
to the understanding of the Einstein--Hilbert theory.''\footnote{E.
Noether
to F. Klein, 29 February 1918,  Nachlass Klein, (SUB), G\"ottingen.}
After this she proceeded to explain where
matters stood with regard
to the key question Klein hoped to answer, namely the relationship
between the classical and relativistic energy equations. Clearly, she
was already deeply immersed in this problem. 

The fundamental results Noether obtained in \cite{Noether1918b} not only provided a general proof of
 Hilbert's Theorem I, they also clarified mathematically how 
 conservation laws arise in  Lagrangian systems 
for
classical mechanics as well as modern field theories.
In her introduction,
she described her approach as one that combined the formal methods of the
calculus of variations with techniques from Sophus Lie's theory of continuous  groups.
Most of Lie's work was motivated by a vision for solving
 general systems of differential equations that admit a given group of
transformations. His pursuit of this program led him to develop what came to be known as the theory of Lie groups.\footnote{For historical
background
on Lie's work and its influence, see
\cite{Haw} and \cite{Hawkins2000}.} Noether pointed out  that within
the context of invariant variational systems one could obtain much stronger theorems than in the general cases handled by Lie.

Noether's ``theorem'' is really two theorems, one dealing with
transformation groups determined by finitely many parameters,
the other concerned with groups determined by finitely
many {\it functions} and their derivatives. Following Lie,
she called the first type a finite continuous group, the second
an infinite continuous group. Of particular significance are those groups
containing both types of structures, which Lie called mixed groups.
With regard to physical interpretations, she noted that
her first theorem generalized the formalism underlying the standard results pertaining to first integrals in classical mechanics, whereas her second theorem constituted ``the most general group-theoretic
generalization of `general relativity''' \cite[240]{Noether1918b}.

She formulated these two theorems as follows:

\begin{quote}

Theorem I. Let $G_{\rho}$ be a finite continuous group with $\rho$ parameters. If the integral
$I$ is invariant with respect to $G_{\rho}$, then $\rho$ linearly independent combinations of the
Lagrangian expressions become divergences, and conversely. The theorem also holds in the
limiting case of infinitely many parameters.
\bigskip

Theorem II. Let $G_{\infty\rho}$ be an infinite continuous group depending on $\rho$ continuous
functions. If the integral
$I$ is invariant with respect to $G_{\infty\rho}$, in which arbitrary functions and their derivatives
up to the $\sigma$th order appear, then $\rho$
identical relations are satisfied between the
Lagrangian expressions and their derivatives up to the order $\sigma$. The converse also holds
here. \cite[238--239]{Noether1918b}

\end{quote}

Theorem I (``Noether's Theorem'') is usually the only result cited in the physics literature. Its importance for physical theories is that it precisely 
 characterizes how  conserved quantities arise from symmetries in variational systems. In a letter to Einstein from 7 January 1926, Noether wrote that ``for me, what mattered in `Invariante Variationsprobleme' was the precise formulation of the 
scope of the principle and, above all, its converse \dots '' \cite[2011: 164]{Kosmann-Schwarzbach2006}.
Likewise, 
 Theorem II characterizes the manner in which identities satisfied by a combination of the Lagrangian expressions and their derivatives come into play.
Hilbert's Theorem I may thus be seen as a special case of Noether's second theorem
corresponding to transformations of the group $G_{\infty 4}$ given by four
functions that depend on the four coordinates of the world--points.

 Noether combined these two key results in order to distinguish between ``proper'' and ``improper''  conservation laws in physics.
Suppose the integral $I$ is invariant with respect to a group $G_{\infty\rho}$. One can then
particularize the functions $p_{\lambda}, \,\lambda = 1,2,\dots ,\rho$ to obtain a {\it finite}
continuous subgroup $G_{\sigma}$ of $G_{\infty\rho}$. The divergence relations
 that arise  will then be fully determined by  this $G_{\sigma}$. Moreover, the divergence relations
associated with $G_{\sigma}$ must, being 
 a subgroup of
$G_{\infty\rho}$,   also be derivable from identities
connecting the Lagrangian expressions and their total derivatives by suitably particularizing the
$p_{\lambda}$.
 Noether called such relations that were derivable from an infinite
 group $G_{\infty\rho}$   improper
(``uneigentliche'') divergence relations; all others were  proper
 (``eigentliche''). 
 From these considerations, she concluded that:
\begin{quote}

The divergence relations corresponding to a finite group $G_{\sigma}$
are improper if and only if $G_{\sigma}$ is a subgroup of an infinite group
with respect to which $I$ is invariant. \cite[254]{Noether1918b}

\end{quote}

As Noether noted, the conservation laws of classical
mechanics as well as those of special relativity theory are proper
 in the above sense. One cannot deduce these as invariants of a
suitably particularized subgroup of an infinite group. In general relativity, on the other hand,
every Lagrangian variational formalism will lead to four identities as a consequence of
the principle of  general covariance. In summarizing these findings,
Noether wrote:

\begin{quote}
Hilbert expressed his assertion regarding the absence of actual energy theorems as
a characteristic attribute of `general relativity theory.' If this assertion is to be literally valid, then the term `general relativity' must be taken more broadly than is usual and extended to groups that depend on $n$
arbitrary functions. \cite[256--257]{Noether1918b}

\end{quote}
From the mathematical standpoint, Noether's analysis provided
a strikingly clear and altogether general answer to the question Klein
had raised about the status of conservation laws in Lagrangian systems.
Her study \cite{Noether1918b} remains today nearly the last word 
 on this subject.\footnote{For remarks on modern refinements of Noether's results, see \cite{Kosmann-Schwarzbach2006}.}

\section{On Hilbert's Revised Theory in \cite{Hilbert1924}}

Klein  took deep satisfaction in the part he was
able to play in elucidating the mathematical underpinnings of
key results in the general theory of relativity. In a letter to Pauli,
he related Einstein's remark about how Klein's third note on
general relativity \cite{Kl-7} had made him ``as happy as a child whose
mother had presented him with a piece of chocolate,'' adding
that ``Einstein is always so gracious in his personal remarks,
in complete contrast to the foolish promotional
efforts (``t\"orichten Reklametum'') undertaken to
honor him.''\footnote{Klein to Pauli, 8 March 1921,
in \cite[79]{Pau-3}.}
 Klein also  made it clear to
Pauli that his  article ``could not pass over Hilbert's
efforts in silence.''\footnote{Klein to Pauli, 8 May 1921,
 in \cite[31]{Pau-3}.} Pauli was skeptical of the various unified field theories that had been advanced by Mie, Weyl, and Einstein \cite[205--206]{Pau-2}. Regarding Mie's theory, he saw no way to deduce the properties of
 the world function  $L$  just by knowing its invariants; there were simply far too many alternatives \cite[189--190]{Pau-2}.
By this time, Hilbert had probably drawn a similar conclusion.

Nevertheless, three years later he  published  a
revised version  of \cite{Hilbert1915} and \cite{Hilbert1917}
in {\it Mathematische
Annalen} \cite{Hilbert1924}. There, he advertised this as ``essentially a reprint
of the earlier communications \dots
and my remarks on them that F. Klein published in \dots 
\cite{Kl-1} -- with only minor editorial alterations and changes in
order to ease understanding'' \cite[1]{Hilbert1924}.
 In truth,
however, this ``reprint'' contains major changes 
 that no careful reader could possibly
miss. These pertain mainly to \cite{Hilbert1915}, the focus of discussion for the present account.

As noted above, Hilbert's invariant energy vector disappears entirely in this revised account. Furthermore, he
softened the physical claim that had been so central for the original theory, namely ``electrodynamic phenomena are the effects of gravitation.'' In place of this, he now wrote that the four independent identities that derive from the gravitational
equations $[\sqrt g H]_{\mu\nu} =0$
signify the ``connection between gravity and
electrodynamics'' \cite[10]{Hilbert1924}. Hilbert noted that his earlier 
 Theorem I had served as the leitmotiv for his theory, but he only mentions it in passing. In a 
 footnote, he cites  Emmy Noether's paper \cite{Noether1918b} for a
``general proof'' \cite[6]{Hilbert1924}. In its place, he refers to a slightly more general version of the result formerly called Theorem III. Here it appears as Theorem 2, but  instead of the expressions (\ref{eq:i_s}) and  (\ref{eq:i_s^l}) Hilbert now writes:
 \begin{equation}\label{Hilbert-Theorem 2}
i_s= \sum_{\mu\nu}( [ \sqrt g J]_{\mu\nu} g^{\mu\nu}_s +  [ \sqrt g J]_{\mu}q_{\mu s})
 \,;\, i_s^l= -2 \sum_{\mu} [ \sqrt g K]_{\mu s} g^{\mu l} +[ \sqrt g J]_{l}q_{s}.
 \end{equation}

As before, (\ref{eq:Hilbert-ident-2}) still holds, and he now applies this theorem successively to $K$ and $L$, following Klein (and implicitly Noether). This leads in the first case  to the four identities (\ref{Klein-identities-1}):
$$ \sum_{\mu\nu}[\sqrt g  K]_{\mu\nu} g^{\mu\nu}_{s}+ 2\sum_{\mu m}\frac{\partial([\sqrt g K]_{\mu s} 
g^{\mu m})}{\partial x_m}
=0, \,\, s = 1,\,2,\,3,\,4.$$
Hilbert also  rewrites the field equations (\ref{eq:Hilbert-grav-eqns})  by  introducing
$$T_{\mu\nu}= -\frac{1}{\sqrt g}\frac{\partial \sqrt g L}{\partial g^{\mu\nu}}.$$
Since  
$[\sqrt g K]_{\mu\nu} = \sqrt g (K_{\mu\nu}-\frac{1}{2}g_{\mu\nu}K)$, the field equations now appear as:
$$K_{\mu\nu}-\frac{1}{2}g_{\mu\nu}K = T_{\mu\nu}.$$ Inserting $L$ in (\ref{eq:Hilbert-ident-2}) yields
 \begin{equation}\label{eq:2-L}
 \sum_{\mu\nu}-(\sqrt g T_{\mu\nu})g^{\mu\nu}_s+2\sum_m \frac{\partial(-\sqrt g T^m_{s})}{\partial x_m} +
\sum_{\mu}[\sqrt g L]_{\mu}q_{\mu s}-\sum_{\mu} \frac{\partial([\sqrt g L]_{\mu}q_s)}{\partial x_{\mu}}
=0.
 \end{equation}
Invoking the field equations $[\sqrt g L]_{\mu}=0$, the last two terms vanish, leaving:
 \begin{equation}
 \sum_{\mu\nu} \sqrt g T_{\mu\nu}g^{\mu\nu}_s+2\sum_m \frac{\partial \sqrt g T^m_{s}}{\partial x_m}=0,
 \end{equation}
which are the familiar equations (\ref{eq:AE-energy3})  for the matter tensor $T_{\mu\nu}$. As Einstein noted in \cite[325]{Ein-6}, these equations are the general relativistic analogue for the classical conservation laws of momentum and energy, where the second term represents the transfer of momentum-energy from the gravitational field to matter.\footnote{This passage is the only place in \cite{Ein-6}  in which Einstein  made direct reference to \cite{Hilbert1915}.} Hilbert remarks accordingly that these equations pass over to true conservation laws when the $g^{\mu\nu}$ are constant, in which case
$$\sum_m \frac{\partial  T^m_{s}}{\partial x_m}=0.$$
Hilbert showed similarly that by invoking the gravitational field equations (\ref{eq:Hilbert-grav-eqns}) the first two terms above vanish, which leaves:
$$\sum_{\mu}[\sqrt g L]_{\mu}q_{\mu s}-\sum_{\mu} \frac{\partial([\sqrt g L]_{\mu}q_s)}{\partial x_{\mu}}
=0.$$
These are four differential equations connecting the electrodynamical Lagrangians with their first derivatives, as asserted by Noether's second theorem. Hilbert's derivation at this key point in \cite{Hilbert1924} follows the argument in \cite{Kl-1} almost to the letter. Thus many of the technical tricks he  employed in \cite{Hilbert1915} have now disappeared, making this paper far easier to follow than the original.

It would seem unlikely that many readers  noticed that \cite{Hilbert1924} was hardly  what could be called 
``essentially a reprint
of the earlier communications.'' Yet even later commentators accepted this characterization at face value, as pointed out in \cite[227]{Rowe1999}. In today's world, with our ready access to so many published sources, one might hope that historians would be held to a higher standard.

 At the outset of his paper, Hilbert noted that only future research could decide  whether a program like the one he first envisioned in 1915 might  actually be realizable. Many physicists would continue to cling  to this 
dream  of establishing a pure field theory that could account for microphysical phenomena, but a growing number had become skeptical. By 1924, Hilbert had begun to immerse himself in the foundational problems of quantum theory, and these would occupy a good part of his attention throughout the 1920s \cite[503--706]{Sauer/Majer2009}. Emmy Noether, on the other hand, would soon emerge to become the leader in G\"ottingen of an important  research school, one whose followers promoted her special vision for abstract algebra. Her venture into mathematical physics, fruitful as it had been, was merely an episode in her early career. If physicists today think of  her in connection with ``Noether's Theorem'' -- by which they mean the first and not the highly significant second theorem in \cite{Noether1918b} -- they typically overlook the role she played in the dramatic, but also highly complex story of how Einstein's theory of gravitation was received in Germany during the years of the First World War.

\end{document}